\documentclass[journal=jacsat,manuscript=article]{achemso}

\usepackage[version=3]{mhchem} 
\usepackage{graphicx}
\usepackage{siunitx}
\usepackage{hyperref}
\usepackage{color}
\usepackage{verbatim}
\usepackage{amssymb}
\usepackage{moresize}
\usepackage{natbib}

\author{Cristina Caruso}
\affiliation[First University]
{Department of Applied Science and Technology, Politecnico di Torino, Corso Duca degli Abruzzi 24, 10129 Torino, Italy}
\author{Martina Crippa}
\affiliation[First University]
{Department of Applied Science and Technology, Politecnico di Torino, Corso Duca degli Abruzzi 24, 10129 Torino, Italy}
\author{Annalisa Cardellini}
\affiliation[Second University]
{Department of Innovative Technologies, University of Applied Sciences and Arts of Southern Switzerland, Polo Universitario Lugano, Campus Est, Via la Santa 1, 6962 Lugano-Viganello, Switzerland}
\author{Matteo Cioni}
\affiliation[First University]
{Department of Applied Science and Technology, Politecnico di Torino, Corso Duca degli Abruzzi 24, 10129 Torino, Italy}
\author{Mattia Perrone}
\affiliation[First University]
{Department of Applied Science and Technology, Politecnico di Torino, Corso Duca degli Abruzzi 24, 10129 Torino, Italy}
\author{Massimo Delle Piane}
\affiliation[First University]
{Department of Applied Science and Technology, Politecnico di Torino, Corso Duca degli Abruzzi 24, 10129 Torino, Italy}
\author{Giovanni M. Pavan}
\affiliation[First University]
{Department of Applied Science and Technology, Politecnico di Torino, Corso Duca degli Abruzzi 24, 10129 Torino, Italy}
\alsoaffiliation[Second University]
{Department of Innovative Technologies, University of Applied Sciences and Arts of Southern Switzerland, Polo Universitario Lugano, Campus Est, Via la Santa 1, 6962 Lugano-Viganello, Switzerland}
\email{giovanni.pavan@polito.it}

\title[An \textsf{achemso} demo]
  {Classification and Spatiotemporal Correlation of Dominant Fluctuations in Complex Dynamical Systems}

\keywords{Fluctuations, Descriptors, Spatiotemporal correlations, Complex molecular systems, Dynamic events}

\begin{document}


\footnotesize
\begin{abstract}
\footnotesize The behavior of many complex systems, from nanostructured materials to animal colonies, is governed by local transitions that, while involving a restricted number of interacting units, may generate collective cascade phenomena. Tracking such local events and understanding how they emerge and propagate throughout these systems represent often a challenge. Common strategies monitor specific parameters, tailored \textit{ad hoc} to describe certain systems, over time. However, such approaches typically require prior knowledge of the underpinning physics and are poorly transferable to different systems. Here we present LEAP, a general, transferable, agnostic analysis approach that can reveal precious information on the physics of a variety of complex dynamical systems simply starting from the trajectory of their constitutive units. Built on a bivariate combination of two abstract descriptors, LENS and \textit{$\tau$}SOAP, the LEAP analysis allows (i) detecting the emergence of local fluctuations in simulation or experimentally-acquired trajectories of any type of multicomponent system, (ii) classifying fluctuations into categories, and (iii) correlating them in space and time. We demonstrate how LEAP, just building on the abstract concepts of local fluctuations and their spatiotemporal correlation, efficiently reveals precious insights on the emergence and propagation of local and collective phenomena in a variety of complex dynamical systems ranging from the atomic- to the microscopic-scale. Given its abstract character, we expect that LEAP will offer an important tool to understand and predict the behavior of systems whose physics is unknown \textit{a priori}, as well as to revisit a variety of known complex physical phenomena under a new perspective.
\end{abstract}

\section{Introduction}
Complex systems, as an ensemble of interacting units, are characterized by non-trivial internal dynamics that are often challenging to unveil. From atomic to macroscopic sizes, global trends often hide local events, fluctuations, or instantaneous perturbations that, although rare, can trigger interesting behaviors.\cite{cavagna2023natural,ballerini2008interaction} A beautiful example is the boson peak in amorphous solids,\cite{nakayama2002boson} that is an excess in the heat capacity whose microscopic origin is attributed to the emergence of dynamical defects or locally ordered structures in uniformly disordered systems.\cite{hu2022origin, gonzalez2023understanding} Local perturbations of the atomic/molecular structure around a critical point are also at the origin of phase transition or nucleation phenomena.\cite{sosso2022role,tao2023data} Material properties, \textit{e.g.} the brittle or ductile macroscopic deformation, are frequently controlled by the emergence and amplification of atomic-level defects.\cite{o2023microscopic} Even on larger scales, the collective behavior observed, for example, in active colloids converting selective energies into motion,\cite{reyes2023magnetic,fasano2019thermally} in bacterial colonies, fish banks, or bird flocks can be controlled by local perturbations or events that involve a restricted number of individuals.\cite{ballerini2008interaction,CAVAGNA2010653,bialek2014social}
Understanding how local fluctuations may emerge, their correlation and amplification in space and time, and how they may determine emergent collective behaviors has therefore important implications in many fields and in the comprehension of complex systems in general. 

Although crucial, localizing rare events and understanding the mechanisms at the root of emergent properties still present some unsolved challenges. Indeed, this requires to (i) unambiguously identify and (ii) classify such fluctuations according to their nature. In recent years, the increased volume of data generated by molecular simulations has led to a growing interest in the development of machine learning (ML)-based methods that are capable of identifying, within high-dimensional datasets, those local events revealing a disruption of structural and dynamical environments in both metals and soft matter.\cite{cardellini2023unsupervised,kashiwa2019phase,arnold2022replacing} Numerous studies and consistent efforts have been focused on tracking local events/fluctuations rising from the time-space evolution of observed variables.\cite{ballerini2008interaction,cavagna2014bird, liu2021activity} Several methods, such as the change point detection,\cite{aminikhanghahi2017survey} eigenvalue spectrum,\cite{trefethen1993hydrodynamic} critical slowing-down,\cite{dakos2008slowing,marconi2020testing} dynamic network biomarkers/markers (DNB),\cite{liu2014early,yang2018dynamic} have been adopted to distinguish critical transition points between two steady-states. Within this framework, the selection of an adequate \textit{descriptor} emerges as a primary task for effectively detecting fluctuations. Successful results have been obtained from system-based descriptors, \textit{i.e.} tailor-made on specific system properties,\cite{steinhardt1981point, stukowski2012structure} thus dependent on prior knowledge about system features. However, the goal is often to retrieve the dynamics of systems whose features are not known \textit{a priori}, whereby general descriptors, \textit{e.g.} based only on the mutual arrangements or movements of neighbor individuals, may show wider applicability.\cite{pietrucci2015systematic, behler2011atom, drautz2019atomic, faber2015crystal, gasparotto2019identifying, musil2021physics} Among them, an advanced representation of atomic environments is provided by the Smooth Overlap of
Atomic Position (SOAP).\cite{bartok2013representing} Coupled with ML approaches, SOAP has enabled the characterization of diverse systems at equilibrium,\cite{monserrat2020liquid, offei2022high, de2016comparing} including soft disordered assemblies\cite{gasparotto2019identifying, cardellini2023unsupervised, capelli2021data, lionello2022supramolecular, gardin2022classifying} and metals.\cite{reinhardt2020predicting, cioni2024sampling, cioni2023innate, rapetti2023machine} In the philosophy of tracking the time-space evolution of local events/fluctuations, two novel general descriptors have been recently developed: Local Environments and Neighbors Shuffling (LENS) \cite{crippa2023detecting} and \textit{Time}SOAP (\textit{$\tau$}SOAP).\cite{caruso2023timesoap} By keeping track of individuals along molecular dynamics (MD) trajectories and any changes within their local environments, \textit{i.e.} neighborhoods, LENS and \textit{$\tau$}SOAP have been demonstrated to carefully characterize the dynamics of a wide range of systems albeit describing different local features.\cite{crippa2023detecting, caruso2023timesoap, cioni2024sampling, becchi2024layer} LENS has been conceived to capture local reshuffling and permutation events that cannot be easily captured with other descriptors (\textit{e.g.}, SOAP).\cite{crippa2023detecting,crippa2023machine} At the same time, LENS cannot capture local structural reconfigurations in the neighborhood of each unit/particle in the system, being a \textit{permutationally variant} and \textit{structurally invariant} descriptor.\cite{crippa2023detecting} On the other hand, \textit{$\tau$}SOAP detects local variations in the order/disorder of the neighbors of every particle in the system. However, while keeping track of local structural reconfigurations, \textit{$\tau$}SOAP cannot capture, \textit{e.g.}, local permutations and reshuffling: \textit{Vice-versa}, it is a \textit{permutationally invariant} and \textit{structurally variant} descriptor.\cite{caruso2023timesoap} 
The complementarity of LENS and \textit{$\tau$}SOAP opens the opportunity to capture different types of local events and study their correlations, unveiling precious information on the physics of a variety of complex systems using an abstract and purely data-driven approach. 

Here, we describe \textbf{LEAP}, an abstract bi-component analysis based on the combination of the \textbf{LE}NS and \textit{$\tau$}SO\textbf{AP} descriptors that allow (i) to detect dominant local fluctuations occurring in complex dynamical systems of any type, (ii) to classify such fluctuations based on their physical nature, and (iii) to correlate them in time and space. Simply starting from experimental or simulation trajectories of the constitutive units/individuals, \textbf{LEAP} can provide unique information on the physics of a variety of complex dynamical systems in an exquisitely agnostic and data-driven manner. We tested \textbf{LEAP} to study various types of complex dynamical systems ranging from the atomic to the microscopic scale. The results that we obtained demonstrate how \textbf{LEAP} can efficiently provide crucial insights useful to understand the mechanisms underpinning a variety of physical phenomena and to trace back them to the spatiotemporal correlations between the local fluctuations that animate the systems.
This provides us with a precious tool and a robust approach to explore and improve our understanding of complex systems whose physics is not known \textit{a priori}, as well as to revisit known physical phenomena under a new light.

\section{Results and discussion}

\subsection{The LEAP Analysis}

In this work, we demonstrate the potential of a \textbf{LEAP} analysis to unravel complex physical phenomena occurring in a range of systems with different inherent dynamical behaviors. As test cases, we study several types of prototypical systems in and out of equilibrium, dominated by rare-local, rather than collective non-local, events. We prove the generality of such an approach, analyzing both simulations and experimentally-collected trajectories ranging from the atomic to the microscopic scales.
 
\begin{figure*}
 \centering
 \includegraphics[width=16cm]{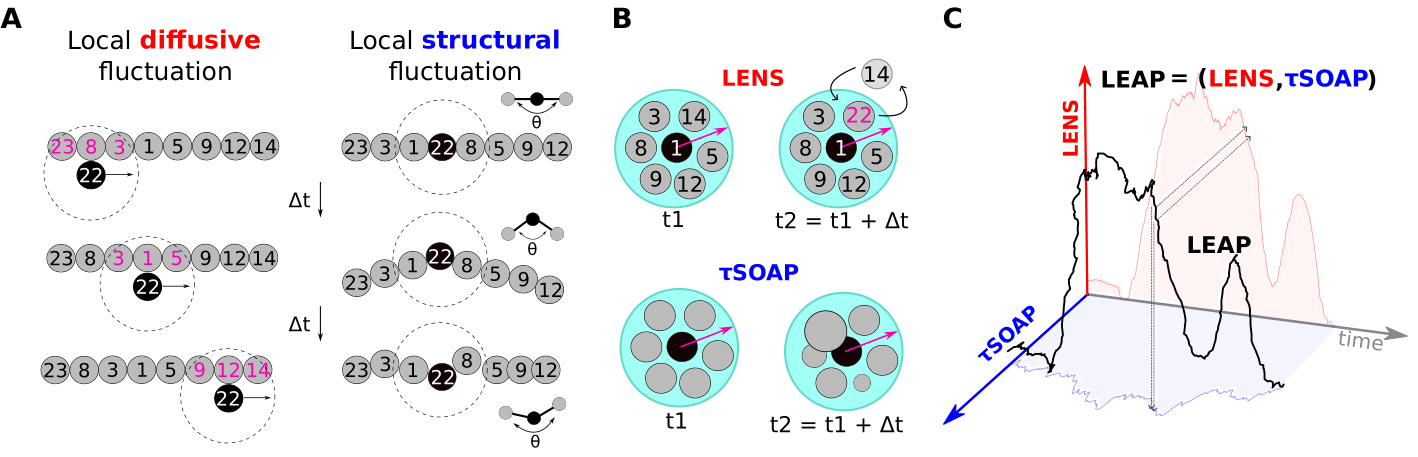}
  \caption{\footnotesize\textbf{LEAP analysis scheme.}(A) Schematic illustration of two types of local events, \textit{i.e.} occurring among individuals (IDs) within a neighborhood (outlined with black dot circles), that can be observed in the trajectories of complex systems. Local \textit{diffusive} fluctuation (Left): ID 22 changes its neighbor's identities yet preserving the structural order in its surrounding. Local \textit{structural} fluctuation (Right): ID 22 undergoes a structural rearrangement without changing the identities of its closest neighbors. (B) Intrinsically distinct insights at the root of LENS and \textit{$\tau$}SOAP molecular descriptors. Over time, and within a certain cutoff neighborhood, LENS (Top) keeps track of changes in each ID's neighbor list, while \textit{$\tau$}SOAP (Bottom) monitors variations in ID's local order. (C) \textbf{LEAP} descriptor (black) as a bivariate time-series joining LENS (red) and \textit{$\tau$}SOAP (blue) components.}
\label{fig:fig01}
\end{figure*}

The philosophy of the analysis relies on considering systems as composed of N interacting individuals (IDs), and on monitoring each of them along the trajectory. We show how such "microscopic" - rather than "macroscopic/average" - approach allows to identify different types of local and collective events and, consequently, to correlate them in space and time thereby providing a thorough description of the inherent physics underlying disparate complex systems. Fig.~\ref{fig:fig01}A shows a representation of two different types of local events that can be observed in the trajectories of complex systems. Considering ID 22 as an example, we report, on the left, an event where the unit is experiencing a "rigid" sliding on the other units, namely a change in its neighbors' identities (highlighted in magenta) without modifying its surrounding structural order: A permutationally variant, though structurally invariant, event. For the sake of simplicity, we refer to this type of event as a "local \textit{diffusive} fluctuation". Differently, on the right, the same unit 22 undergoes a structural rearrangement of its closest neighbors, which yet preserve the same identities: \textit{Vice-versa}, a structurally variant but permutationally invariant event. This is an example of what we mean, for ease, with the term "local \textit{structural} fluctuation". Noteworthy, in realistic systems, these two different types of events are not easy to capture, nor it is easy to distinguish between them. Such fundamental distinctions are relevant in the perspective of describing higher-scale dynamical events occurring in the systems.  

Recently, LENS \cite{crippa2023detecting} and \textit{$\tau$}SOAP \cite{caruso2023timesoap} molecular descriptors have demonstrated to accurately detect these local dynamical events described above (see \textit{Materials and Methods} for details). As illustrated in Fig.~\ref{fig:fig01}B, LENS and \textit{$\tau$}SOAP are perfectly tailored to capture purely \textit{diffusive} (Top) and purely \textit{structural} (Bottom) fluctuations, respectively. Although conceived to be different, these two novel descriptors are, at the same time, perfectly complementary to each other.
Motivated by these observations, here we demonstrate how combining LENS and \textit{$\tau$}SOAP into a \textbf{LEAP} analysis allows to get a unique characterization of distinct systems. 

An illustration of \textbf{LEAP} (black), defined as $\textbf{LEAP} = (LENS, \textit{$\tau$}SOAP)$, namely a bivariate time-series composed of LENS (red) and \textit{$\tau$}SOAP (blue) components, is reported in Fig.~\ref{fig:fig01}C. Noteworthy, both LENS and \textit{$\tau$}SOAP have been normalized from 0 to 1 in the \textbf{LEAP} definition (additional details are provided in the \textit{Materials and Methods} section).
In the following, we prove the broad applicability of our analysis not only in classifying local events occurring in the system into different types of local fluctuations but also in correlating such different fluctuations in space and time. This allows unveiling the overarching behavior in several prototypical systems dominated by diverse inherent physics with a growing level of internal complexity.

\subsection{LEAP fluctuations at the Ice/liquid water equilibrium interface}
\begin{figure*}
 \centering
 \includegraphics[width=16cm]{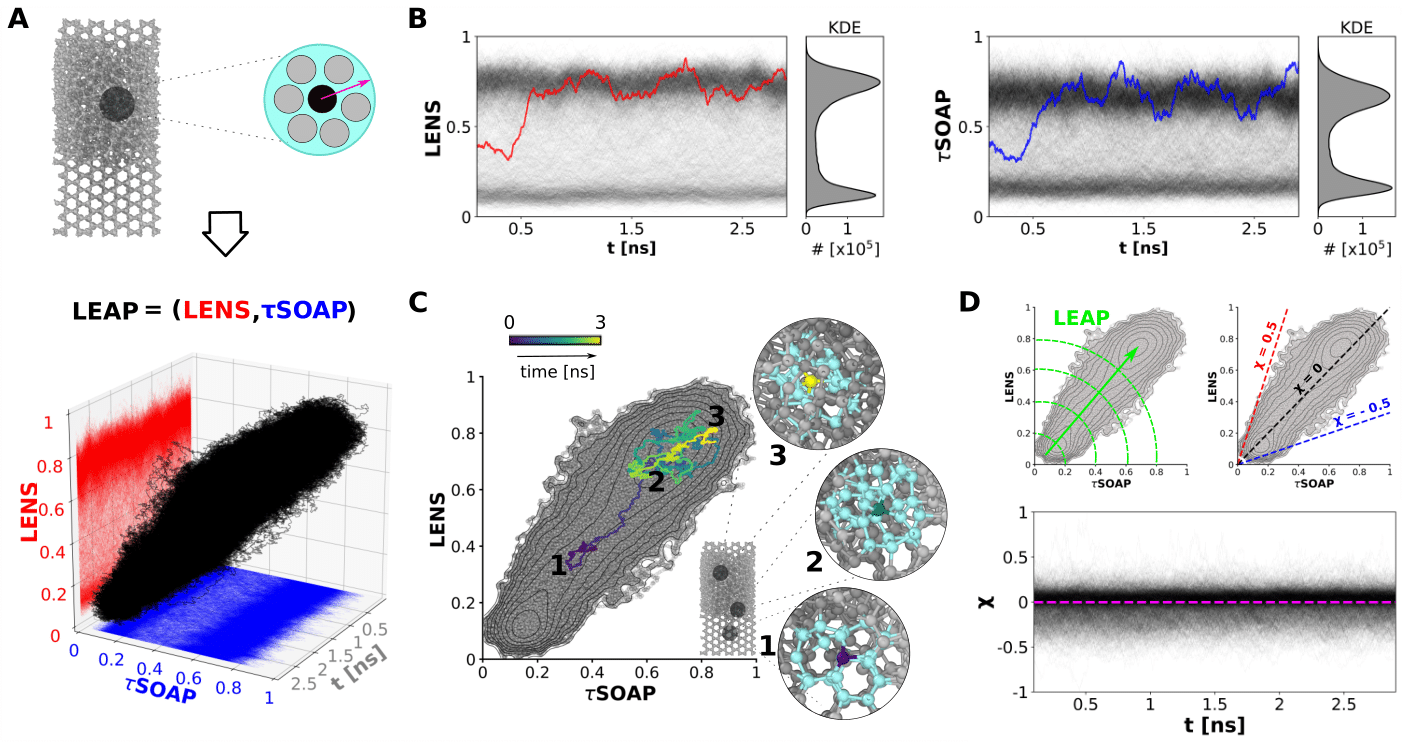}
  \caption{\footnotesize\textbf{LEAP analysis in Ice/liquid water phase coexistence.} (A) \textbf{LEAP} time-series dataset related to 3 ns extracted from a 100 ns-long MD trajectory (from 95 ns to 98 ns, sampling time $\Delta$t = 0.001 ns) composed of 2048 water molecules (\textbf{TIP4P/Ice} water model),\cite{abascal2005potential} whose 50\% arranged in the crystalline hexagonal ice configuration and the remaining 50\% in the liquid phase, coexisting in a dynamic equilibrium. Oxygen atoms (OW) are considered as centers to compute both LENS and \textit{$\tau$}SOAP. (B) LENS and \textit{$\tau$}SOAP time-series, with the related Kernel Density Estimation (KDE) distributions, for all the water molecules in the system. Signals related to an example water molecule (ID 595) are highlighted on both LENS (red) and \textit{$\tau$}SOAP (blue) components. (C) Projection of the whole \textbf{LEAP} dataset on the 2D LENS-\textit{$\tau$}SOAP phase space (3000 time steps for 2048 water molecules, for a total of $\sim$ \num{6e6} data points displayed in gray). The \textbf{LEAP} path related to the ID 595 is colored from blue to yellow as time increases. An analogous color code is used for the example water molecule (ID 595) in the representative MD snapshots taken at t = 0 ns (1), t $\sim$ 0.5 ns (2) and t $\sim$ 2.8 ns (3), while its neighborhood is colored in cyan. (D) $\chi$ parameter. Top: \textbf{LEAP} magnitude (Left) and $\chi$ (Right), represented on the 2D \textbf{LEAP} phase space. Bottom: $\chi$ parameter over time is reported for all the water molecules in the system, with the mean value ($\chi$ $\sim$ 0) highlighted in magenta.} 
\label{fig:fig02}
\end{figure*}

As a first case study, we show the results of a \textbf{LEAP} analysis on the trajectories of water molecules in a system where ice and liquid water coexist in a dynamic equilibrium (Fig.~\ref{fig:fig02}). 

Fig.~\ref{fig:fig02}A (Top) shows an MD periodic simulation box containing 2048 \textbf{TIP4P/Ice} water molecules, of which 50\% are arranged in the crystalline hexagonal ice configuration and the other 50\% are in the liquid phase. This system is simulated for 100 ns, using a sampling time interval of $\Delta$t = 0.001 ns, in correspondence of the solid/liquid transition temperature for the employed \textbf{TIP4P/Ice} water model.\cite{abascal2005potential, garcia2006melting} Herein, we focus on 3 ns, extracted from the last part of the 100 ns-long MD simulation, and we compute the bivariate \textbf{LEAP} time-series for each water molecule (see Methods for computational details). The \textbf{LEAP} time-series dataset, related to all the water molecules in the system, clearly shows a \textit{linear} trend (Fig.~\ref{fig:fig02}A, Bottom).
This means that the \textit{diffusive} fluctuations -detected by LENS- and the \textit{structural} fluctuations -captured by \textit{$\tau$}SOAP- concurrently occur.
Calculating the LENS and \textit{$\tau$}SOAP time-series for each ID in the system, similar Kernel Density Estimation (KDE) distributions can be observed (Fig.~\ref{fig:fig02}B): Two peaks, at $\sim$ 0.2 and at $\sim$ 0.8 intensity, emerge in both signals. Therefore, two main domains presenting clear dynamic fingerprints can be identified in the system. As a result, an univariate Onion clustering analysis\cite{becchi2024layer} carried out separately on LENS and \textit{$\tau$}SOAP KDEs (see SI Appendix, Fig. S1) detects, in both cases, ice (lowest descriptors' intensities), liquid water (highest descriptors' intensity), and solid-liquid interface (intermediate values). A detail of the LENS and \textit{$\tau$}SOAP signals related to an example water molecule, ID 595, is highlighted on each plot in red and blue, respectively. The marked signals display a clear correspondence between the two time-series: At t $\sim$ 0.5 ns, the ID 595 simultaneously undergoes a LENS and \textit{$\tau$}SOAP transition. Confirming our previous findings,\cite{caruso2023timesoap, crippa2023detecting, crippa2023machine} the correspondence of the identified domains is a first evidence that both LENS and \textit{$\tau$}SOAP descriptors come up to be equivalent in identifying water states in phase coexistence. 

The projection of the \textbf{LEAP} time-series onto its 2D LENS-\textit{$\tau$}SOAP phase space is shown in Fig.~\ref{fig:fig02}C for each molecule in the system. In such 2D plot, each single point displayed in gray is the \textbf{LEAP} value related to a specific water molecule in a precise MD time step (3000 time steps for 2048 water molecules, for a total of $\sim$\num{6e6} data points). Projected on this plot, the \textbf{LEAP} path of the ID 595, chosen as a representative example, is colored from blue to yellow as time increases. The same water molecule (shown with the analogous color code, together with its neighborhood in light cyan) is depicted in the three MD snapshots in Fig.~\ref{fig:fig02}C while diffusing from the ice (1) to the liquid phase (3). During the transition, the water molecule moves on the 2D phase space along the diagonal meaning that, in this case, to every local structural rearrangement (\textit{$\tau$}SOAP) corresponds an equivalent change in the local reshuffling dynamics (LENS). Similar results have been found by analyzing a longer (50 ns) MD trajectory, corresponding to the whole second half of such 100 ns-long MD simulation (see Fig. S2 in the SI Appendix). 

The pathway that molecules' trajectories follow onto such LENS-\textit{$\tau$}SOAP phase space clearly determines the type of correlation between \textit{diffusive} and \textit{structural} fluctuations. To quantitatively describe this correlation, we define the following parameter: 
\begin{equation}
\centering
    \chi=\frac{LENS - \textit{$\tau$}SOAP}{LENS + \textit{$\tau$}SOAP}.
\label{eq:one}
\end{equation}
Over time, $\chi$ allows to measure to what extent the system's dynamics is driven by a specific type of fluctuation. When $\chi$ $\sim$ 0, LENS and \textit{$\tau$}SOAP changes simultaneously occur. On the other hand, $\chi$ $<$ 0 and $\chi$ $>$ 0 indicates that the unit ID is interested by, respectively, \textit{$\tau$}SOAP-dominated and LENS-dominated fluctuations.

In such a specific case (the simplest we investigate, where LENS and \textit{$\tau$}SOAP provide equivalent insights), the average $\chi$ is roughly 0, as displayed in magenta in Fig.~\ref{fig:fig02}D. This points out the correspondence between fluctuations in the structural order and in the neighbor identities while molecules move across phase transitions. However, there are systems exhibiting non-trivial behaviors which convey in non-simultaneous \textit{diffusive} and \textit{structural} fluctuations and, hence, in a peculiar LENS-\textit{$\tau$}SOAP mismatch, where further investigations are needed.  

\subsection{Non-trivial dynamical fluctuations on an atomic metal surface}
\begin{figure*}
 \includegraphics[width=16cm]{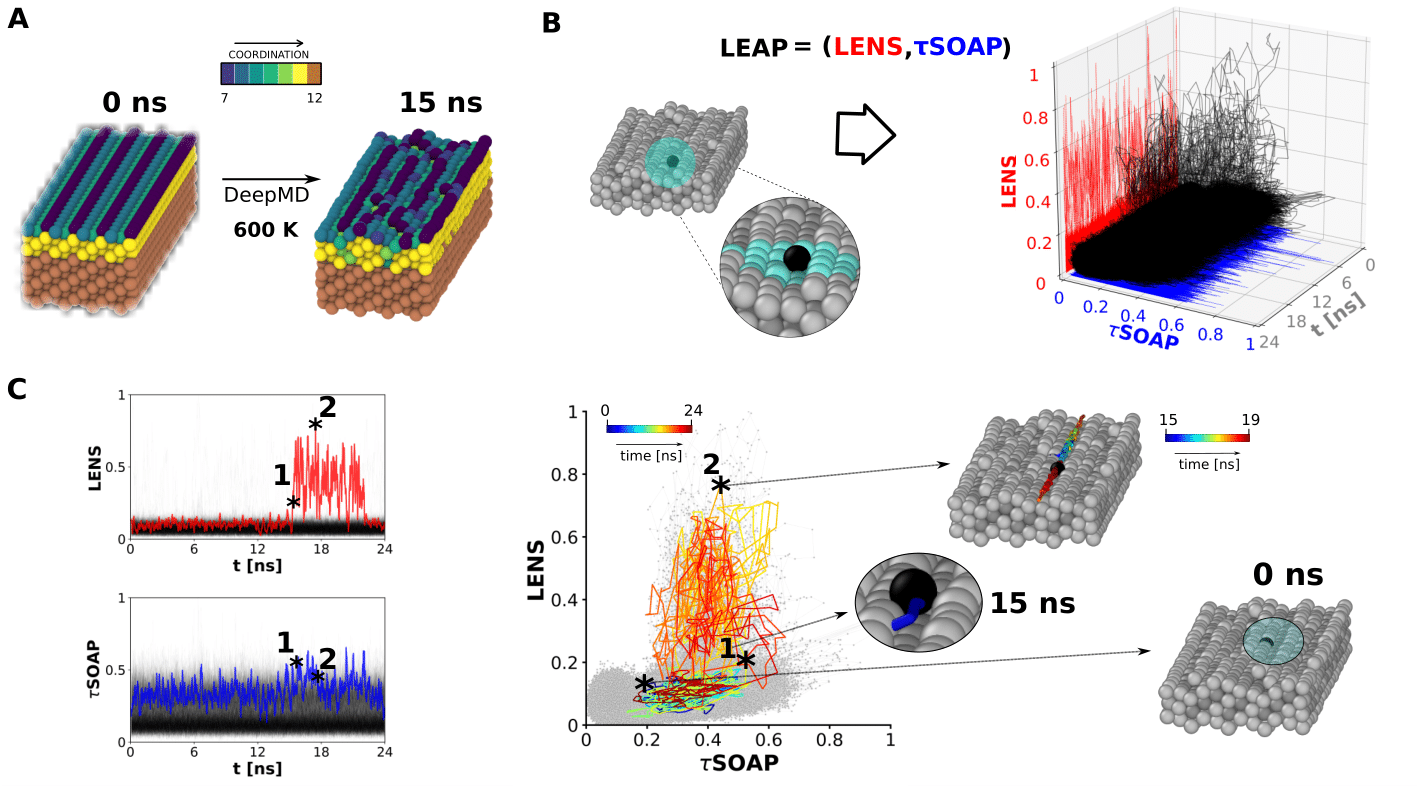}
 \centering
  \caption{\footnotesize\textbf{LEAP fluctuations in a metal surface dominated by rare events.} (A) Portion of \textbf{Cu(211)} face-centered cubic (FCC) surface, composed of 2400 atoms, simulated by DeepMD-based potential at T = 600 K. Two snapshots, taken at t = 0 ns and t = 15 ns, are colored according to their coordination number. (B) \textbf{LEAP} time-series dataset related to the three top-most layers (995 atoms) of the \textbf{Cu(211)} FCC surface. Left: Focus on the local neighborhood representation of a central Cu atom (black) and its closest neighbors (cyan). Right: LENS (red), \textit{$\tau$}SOAP (blue), and \textbf{LEAP} (black) time-series plotted along the considered 24 ns of the MD trajectory. (C) Left: LENS and \textit{$\tau$}SOAP time-series, corresponding to all the 995 Cu atoms, are reported in black. The LENS and \textit{$\tau$}SOAP signals of an example Cu atom (ID 460) are highlighted in red and blue, respectively. Right: 2D \textbf{LEAP} dataset related to the Cu atoms on the three top-most layers (995 atoms for 2000 time steps, for a total of $\sim$ \num{2e6} data points displayed in gray). Peculiar trajectory path, related to the example Cu atom (ID 460), colored from blue to red as time increases. Three MD snapshots taken at t = 0 ns (atom well incorporated in the surface), t = 15 ns (point (1), atom detaching from the surface) and 15 ns $<$ t $<$  19 ns (point (2), sliding after the detachment), are reported.}
\label{fig:fig03}
\end{figure*}

As a second case study, we show the results of a \textbf{LEAP} analysis related to a metal copper (Cu) surface which exhibits a peculiar dynamics even well below the melting temperature.\cite{spencer1986stable, jayanthi1985surface} The system consists of a portion of \textbf{Cu(211)} face-centered cubic (FCC) surface, containing 2400 atoms, simulated for 150 ns using a DeepMD-based potential\cite{wang2018deepmd} which allows to reach length and time scales typically not affordable to DFT calculations.\cite{cioni2023innate} In particular, this \textbf{Cu(211)} FCC slab has been simulated at T = 600 K (T$\sim$1/3 of the Cu melting temperature, see \textit{Materials and Methods} for simulation details), thus near the H\"{u}ttig temperature where metal surface dynamics can be observed (Fig.~\ref{fig:fig03}A). Recently, LENS-based analyses have highlighted, for the first time, the presence of sparse atoms sliding in a "rigid" manner on this surface.\cite{crippa2023detecting, crippa2023machine} Interestingly, such fascinating events had not been captured before by either looking at the average system's properties or using traditional pattern recognition analyses based on structural descriptors, \textit{e.g.} SOAP.\cite{cioni2023innate} This is thus an ideal, prototypical example system where local (purely \textit{diffusive}) fluctuations emerge only (predominantly) in the LENS dimension, while no evident fluctuations can be observed in the \textit{$\tau$}SOAP one.

Herein, we consider the three top-most layers (995 atoms) of the \textbf{Cu(211)} FCC slab described above (see the snapshot in Fig.~\ref{fig:fig03}B). For each Cu atom considered, we analyze 24 ns extracted from the last part of the 150 ns-long MD simulation, sampled every $\Delta$t = 12 ps. In Fig.~\ref{fig:fig03}B (Right) we report the whole time-series dataset including the 995 LENS, \textit{$\tau$}SOAP, and \textbf{LEAP} signals plotted in red, blue, and black, respectively. Contrary to the water case of Fig.~\ref{fig:fig02}, here the \textbf{LEAP} dataset is far from being linear in the LENS-\textit{$\tau$}SOAP space. Such a peculiar non-linearity is investigated by first decoupling the \textbf{LEAP} time-series in its LENS and \textit{$\tau$}SOAP components (see Fig.~\ref{fig:fig03}C, Left). Clearly, the two descriptors display diverse trends, as also emphasized by the distinct fingerprints of LENS (in red, Top-Left) and \textit{$\tau$}SOAP (in blue, Bottom-Left) corresponding to the same representative atom (ID 460). Indeed, while LENS deviates from the average behavior after $\sim$ 15 ns, \textit{$\tau$}SOAP slightly enhances. This represents an evident example where the two types of local events (diffusive and structural) do not simultaneously occur. 

To deepen this concept, in Fig.~\ref{fig:fig03}C (Right) we plot the whole \textbf{LEAP} data set projected onto LENS-\textit{$\tau$}SOAP phase space (995 Cu atoms for 2000 time steps, for a total of $\sim$\num{2e6} data points displayed in gray) and we select the peculiar path corresponding to the Cu atom ID 460. Coloring the selected trajectory from blue to red as a function of simulation time, we observe that during the early MD steps the atom path lies on a domain characterized by LENS $<$ 0.2 and \textit{$\tau$}SOAP $\sim$ 0.2, which describes the atom incorporated (and vibrating) in the top-most surface layer (see the MD snapshot at $t = 0$ ns). At roughly 15 ns, the \textbf{LEAP} values increase dramatically: Denoted by (1) in Fig.~\ref{fig:fig03}C, the \textit{$\tau$}SOAP signal increases up to $\sim$ 0.5, while LENS remains $\sim$ 0.2. This indicates a first structural transition, namely the detachment of the atom from the surface (MD snapshot at 15 ns in Fig.~\ref{fig:fig03}C). From 15 to 19 ns, data shows a sharp enhancement of the LENS component, while the \textit{$\tau$}SOAP value remains substantially constant. Denoted by (2), we point out the atom sliding along the surface, after the detachment occurring at $t = 15$ ns. In such a "rigid" event, the sliding atom continuously changes the IDs of its neighbors (the LENS signal reaches $\sim$ 0.8) while the structural order in the surrounding atoms exhibits only slight variations (sliding trajectory line displayed in the snapshot (2), colored according to the time evolution from 15 to 19 ns). After $\sim$ 20 ns of MD, the atom comes back to the starting phase space domain (dark red line, overlapped to the dark blue one), meaning that it is stably reincorporated in the surface. This example reveals rare atomic events occurring on a metal surface at T = 600 K, which have a characteristic physics. Such an atomic surface diffusion is, in fact, characterized by sharp LENS signals, not accompanied by appreciable \textit{$\tau$}SOAP fluctuations. At the same time, the trajectory analysis reveals that a sliding motion is conditioned by a preliminary structural transition, namely a sharp \textit{$\tau$}SOAP variation with negligible LENS change (1). Interestingly, a sharp \textit{$\tau$}SOAP variation occurs both when the atom jumps out and when it is reincorporated in the surface. The sliding mechanism follows a \textit{$\tau$}SOAP$\xrightarrow{}$LENS$\xrightarrow{}$\textit{$\tau$}SOAP sequence. As reported in SI Appendix (Fig. S3), predominantly LENS or predominantly \textit{$\tau$}SOAP fluctuations may emerge also for other atom ID trajectories in this system. 

In cases like this, where the internal dynamical events are characterized by non-simultaneous LENS and \textit{$\tau$}SOAP fluctuations, the sequence and localization of events might be crucial to investigate the physics of the system. This opens, indeed, the possibility to understand (i) if there is a correlation between the number of structural fluctuations needed to generate diffusive LENS events (or \textit{vice-versa}), and (ii) how spatially correlated they have to be in such a way to generate a dynamical transition in the material.   

\begin{figure*}
 \centering
 \includegraphics[width=16cm]{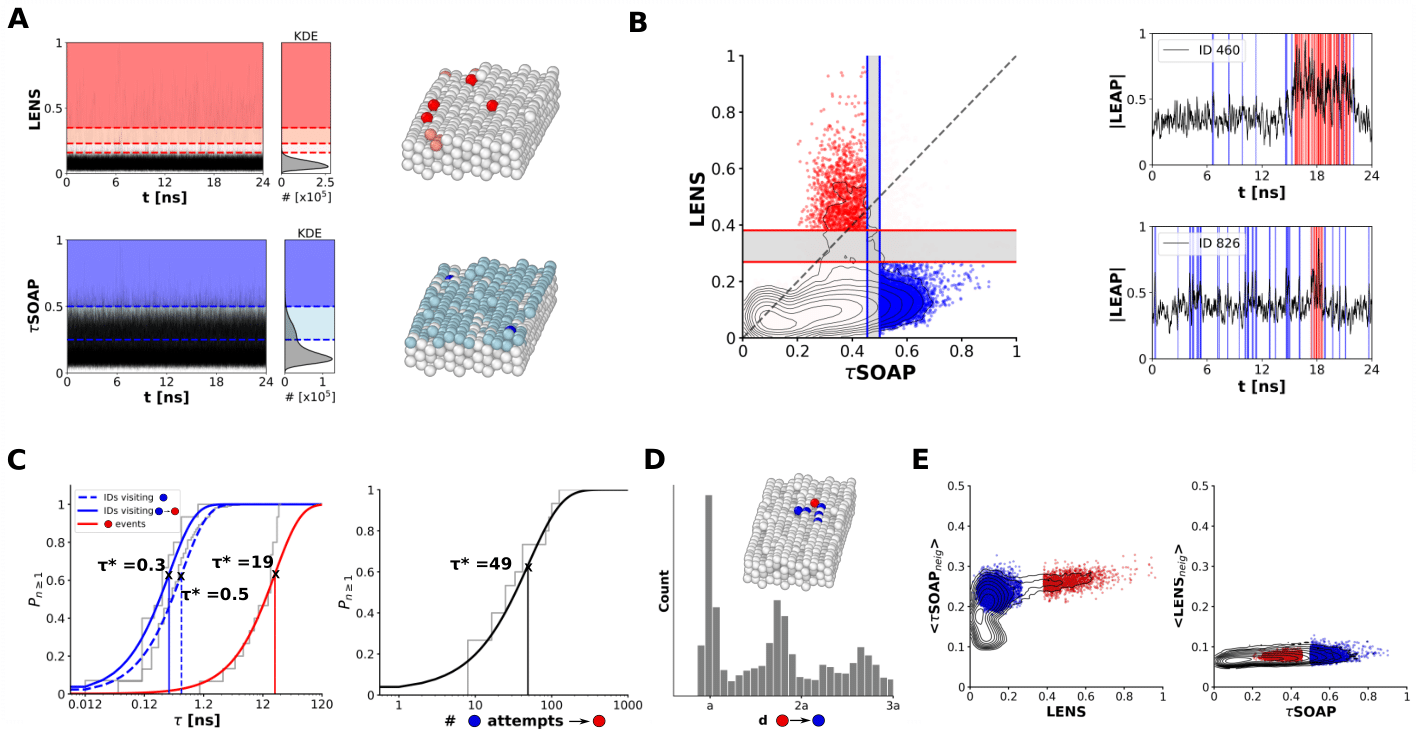}
  \caption{\footnotesize\textbf{Time-space correlation between diffusive and structural fluctuations using \textbf{LEAP}.} (A) LENS and \textit{$\tau$}SOAP time-series, computed for the three top-most layers (995 Cu atoms) of the \textbf{Cu(211)} FCC surface, with the corresponding KDE distributions. Univariate Onion clustering\cite{becchi2024layer} is applied. Clusters are colored from white to red (LENS, Top) or from white to blue (\textit{$\tau$}SOAP, Bottom). The MD snapshots reported on the right, taken at the same time step (t $\sim$ 11 ns), are colored according to the identified clusters, respectively. (B) Left: Projection of the \textbf{LEAP} dataset on the 2D LENS-\textit{$\tau$}SOAP phase space. LENS and \textit{$\tau$}SOAP outlier domains are colored in red and blue, respectively. Right: Time-series of the \textbf{LEAP} magnitude (black) for two example IDs. The transit through LENS and \textit{$\tau$}SOAP outlier domains (red and blue regions shown on the left) are identified with red and blue bands, respectively. (C) Characteristic time scale ($\tau*$) estimation for LENS and \textit{$\tau$}SOAP fluctuations. Left: Cumulative distribution functions (CDFs) for the mean time interval ($\tau$) between successive \textit{$\tau$}SOAP fluctuations related to (i) trajectories visiting the blue region in (B) (dashed blue line), (ii) the subset of (i) crossing the red region in (B) after the blue one (solid blue line); (iii) CDF related to the time interval between LENS fluctuations occurring in the system (red line). Right: CDF for the number of \textit{$\tau$}SOAP fluctuations preceding a LENS fluctuation. (D) Space correlation between LENS and \textit{$\tau$}SOAP fluctuations. \textit{x}-axis: Distance between atoms lying in the LENS outlier region and atoms in the \textit{$\tau$}SOAP outlier region, where (a) is the interatomic distance (a $\sim$ 3.6 \r{A} in FCC Cu). (E) Left: ID LENS value \textit{vs.} \textit{$\tau$}SOAP mean value of its neighbors. Right: ID \textit{$\tau$}SOAP value \textit{vs.} LENS mean value of its neighbors. Blue and red points correspond, respectively, to the blue and red domains in (B).}
\label{fig:fig04}
\end{figure*}

\subsection*{Space $\&$ time correlations between different fluctuation types}
A more detailed analysis of the dynamics of the \textbf{Cu(211)} FCC surface can be obtained by (i) classifying the local fluctuations emerging in the system into structural (\textit{$\tau$}SOAP) and diffusive (LENS) fluctuations, and (ii) analyzing their correlation in space and time. First, in Fig.~\ref{fig:fig04}A we report the LENS and \textit{$\tau$}SOAP time-series with the corresponding KDE distributions. An univariate Onion clustering analysis\cite{becchi2024layer} of the individual LENS and \textit{$\tau$}SOAP KDEs identifies \textit{different} clusters in the two components, displayed in white, light red/blue, and dark red/blue colors (Fig.~\ref{fig:fig04}A). Except for the crystalline bulk atoms characterized by the lowest descriptors' values (first KDE peaks and white atoms in both MD snapshots), the other domains do not match each other. For example, the two clusters corresponding to the highest LENS and \textit{$\tau$}SOAP values (dark red and blue, respectively) do not include the same atoms (compare the red and blue atoms in the MD snapshots in Fig.~\ref{fig:fig04}A, Right). While the \textit{$\tau$}SOAP intermediate cluster (0.25 $\leqslant$ \textit{$\tau$}SOAP $<$ 0.5) well identifies the top-most surface layer, this is not the same in the LENS component. This indicates that the LENS and \textit{$\tau$}SOAP descriptors detect different types of dynamical events on this metal surface. Fig.~\ref{fig:fig04}B, indeed, shows that dominant LENS and \textit{$\tau$}SOAP fluctuations are well distinguishable and detectable as two distinct and separated domains in the \textbf{LEAP} 2D space. Projected on such phase space, the data points that are classified with a confidence interval $>$ 95\% as LENS outliers are colored in red (top-most red cluster in Fig.~\ref{fig:fig04}A), while the data points identified as \textit{$\tau$}SOAP outliers with a confidence interval $>$ 95\% are shown in blue (top-most blue cluster in Fig.~\ref{fig:fig04}A). 

It is interesting to look into the temporal correlations between LENS (red) and \textit{$\tau$}SOAP (blue) fluctuations. Going back to the \textbf{LEAP} time-series, this classification (\textit{$\tau$}SOAP \textit{vs.} LENS fluctuations) allows us to color atoms based on the dynamical event they are experiencing over time (predominantly \textit{$\tau$}SOAP, predominantly LENS or simultaneously occurring). On the right side of Fig.~\ref{fig:fig04}B, two example \textbf{LEAP} time-series (magnitude) for two representative atoms are reported in black: LENS and \textit{$\tau$}SOAP (outliers) fluctuations are identified in red and blue bands, respectively. As clearly shown in both plots (see also SI Appendix, Fig. S4), the selected IDs experience several structural fluctuations (sequence of blue bands) before sliding on the metal surface (red bands). Extending this analysis to all the atoms visiting the outlier (relevant fluctuation) regions, we can obtain a quantitative characteristic time estimation for the LENS and \textit{$\tau$}SOAP fluctuations occurring in the system. For each atom crossing the \textit{$\tau$}SOAP fluctuation region (blue domain in Fig.~\ref{fig:fig04}B), we compute the mean time interval ($\tau$) between two successive \textit{$\tau$}SOAP events, thereby obtaining the cumulative distribution functions (CDFs) shown in Fig.~\ref{fig:fig04}C (see \textit{Materials and Methods} for technical details). The characteristic transition times ($\tau*$, Fig.~\ref{fig:fig04}C, Left) of the structural re-ordering are reported by distinguishing atom trajectories visiting the \textit{$\tau$}SOAP outlier region (dashed blue line) from their subset also experiencing the sliding event (solid blue line). In addition, the characteristic curve related to the sliding events is displayed (in red). It is worth noting that the probability to observe \textit{$\tau$}SOAP (blue) or LENS (red) fluctuations in the system follows the typical Poisson distribution, which is expected for rare transition events. In detail, Fig.~\ref{fig:fig04}C (Left) confirms the rare nature of sliding events which, indeed, happen less frequently than the structural re-ordering: 19 ns and $0.5$ ns are the estimated characteristic times of sliding and rearrangement before sliding, respectively. Furthermore, the blue CDFs reveal that the characteristic transition time scale of structural fluctuations is slightly shorter when atom's rigid surface motion (LENS fluctuation) follows after ($\tau*$ = 0.3 ns for the solid blue, while $\tau*$ = 0.5 ns for the dashed blue line). This shows how, in this system, a local diffusive LENS event is always anticipated by a set of blue fluctuations occurring with a frequency that is $\sim65-70 \%$ higher than the normal one (namely, compared to the average frequency observed for blue fluctuations in general in the system). Fig.~\ref{fig:fig04}C (Right) also shows the number ($\sim49$) of \textit{$\tau$}SOAP structural re-ordering transitions preceding the observation of every atomic sliding (LENS event) on the surface. This indicates that in a surface where structural fluctuations are ubiquitous, observing one rigid atomic sliding on the surface is a rare event (see also \textit{Materials and Methods}) that, statistically, is observed when one atom undergoes a set of \textit{$\tau$}SOAP that are considerably more packed in time (more frequent) than what happens on average.

This data provides a mechanistic picture in terms of temporal correlation between the different types of fluctuations characterizing such a system. As a next step, we also investigate the spatial correlation between them. Fig.~\ref{fig:fig04}D reports the probability density to observe structural events within a certain distance from diffusive events. Data shows the presence of clear peaks, with the first one in correspondence of the inter-atomic distance (a $\sim$ 3.6 \r{A} in FCC Cu). This means that atoms, during the sliding motion, are close to IDs which undergo a reconstruction transition. A closer look at this phenomenon is provided in the Movie S1. For each ID in each MD time step, the plot of Fig.~\ref{fig:fig04}E shows (on the left) the relationship between the ID LENS value and the \textit{$\tau$}SOAP mean value of its neighbors, and \textit{vice-versa} (on the right), between the ID \textit{$\tau$}SOAP value and the LENS mean value of its neighbors (additional correlations are reported in SI Appendix, Fig. S5). The two distributions show how, while LENS fluctuations require, in general, neighbors' \textit{$\tau$}SOAP values higher than the average, \textit{$\tau$}SOAP fluctuations do not necessarily need high neighbors' LENS values. Altogether, this data demonstrates how the sliding events identified as LENS fluctuations are (i) anticipated in time by structural (\textit{$\tau$}SOAP) reconfigurations being, in intensity and frequency, higher than the average and (ii) characterized in space by local rearrangements of close neighborhoods.

The data reported herein demonstrates how classifying fluctuations into different categories and correlating them in space and time hold a great potential in revealing the physics underpinning complex system. This is particularly relevant for characterizing the behavior of systems whose underlying physics is unknown, but also to revisit well-established physical phenomena.

\subsection{Plastic deformation of metals seen under the LEAP analysis}
To prove the generality of the \textbf{LEAP} analysis, we tested it to revisit a well-known phenomenon controlled by the formation and amplification of local defects/fluctuations, namely the deformation and fracture of metals under tensile stress.

\begin{figure*}
 \centering
 \includegraphics[width=16cm]{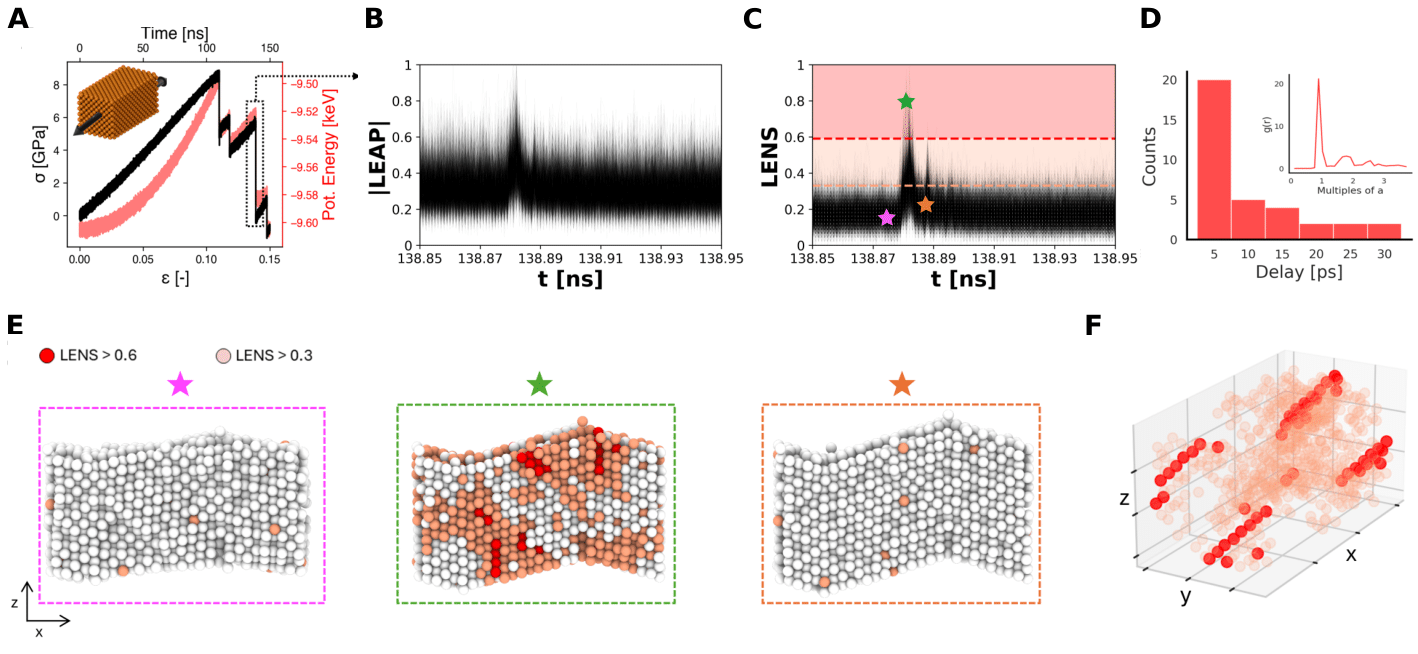}
  \caption{\footnotesize\textbf{LEAP analysis of bulk metal (Cu) during constant strain rate at T = 300 K.} (A) Stress-strain curve (black) over time with corresponding potential energy profile (red), indicating the regions of elastic and plastic deformation. Inset showing the simulated bulk copper structure, with arrows indicating the direction of strain. (B) Time-series of the LEAP signal identifying one detail of the plastic phase highlighted in panel (A). (C) Time-series of the LENS signal within the same timeframe, color-coded by intensity thresholds (LENS $>$ 0.3 in pink, LENS $>$ 0.6 in red); stars mark snapshots detailed in (E). This panel reveals how changes in LENS values dominate the LEAP signal over time. (D) Temporal and spatial correlation analysis of high-intensity LENS signals, showing that most LENS spikes occur within 5 ps of each other and tend to cluster spatially (the inset shows the radial distribution function, g(r), of atoms with high LENS signals, illustrating their spatial proximity). (E) Snapshots corresponding to the marked events in (C), color-coded to show the intensity of their LENS signals, highlighting both spatial and temporal correlations. (F) Three-dimensional representation of atomic positions with high LENS signals, emphasizing regions of coherent plastic deformation, identified as dislocation lines.}
\label{fig:fig05}
\end{figure*}

As a reference case, we consider a bulk of copper (Cu) FCC crystal containing 2744 atoms and subjected to a constant strain rate at $T = 300$ K. Fig.~\ref{fig:fig05}A shows the stress-strain curve alongside the potential energy profile obtained from the MD simulation of a periodic box using an Embedded Atom Method (EAM) potential\cite{mishin2001structural}(see \textit{Materials and Methods} section for details). The plot clearly shows the transition from the elastic to the plastic deformation phase (after $\sim$ 110 ns of MD and $\epsilon$ $>$ 0.11). We use the \textbf{LEAP} analysis to gain insights into one reference plastic event (occurring at $\sim$ 130-145 ns, as highlighted in Fig.~\ref{fig:fig05}A).
Fig.~\ref{fig:fig05}B shows a zoom in the temporal evolution of the \textbf{LEAP} (magnitude) time-series, revealing a pronounced peak at $\sim$ 138.88 ns. Interestingly, the decomposition of such \textbf{LEAP} signal in its LENS (Fig.~\ref{fig:fig05}C) and \textit{$\tau$}SOAP (SI Appendix, Fig. S6) components demonstrates that such \textbf{LEAP} fluctuation is governed by the LENS component, that is, by diffusive events (neighbors reshuffling) rather than by structural reconfigurations. This is consistent with the traditional view of plastic deformations in metals that proceed predominantly through dislocation planes, as also supported by the findings in Fig.~\ref{fig:fig05}D, where temporal and spatial correlation analyses are reported. The data demonstrates the collective and concerted nature of the fluctuations that control the deformation of the material under tensile stress after entering the plastic region. In particular, the plastic deformation, controlled by the motion of the dislocation planes respect to each other, is reflected by sharp LENS fluctuations that are substantially simultaneous in time and correlated in space: They occur within $\sim$ 5 ps (see Fig.~\ref{fig:fig05}D) and with a distance corresponding to the inter-atomic spacing (a), as shown in the inset of Fig.~\ref{fig:fig05}D. Fig.~\ref{fig:fig05}E shows that the dislocation first nucleates in the system and, when the nuclei grow, collective motions of planes emerge (identified by the red domains in the central green rectangle); Then, after the sliding of the dislocation planes, the system comes back and rests. Lastly, Fig.~\ref{fig:fig05}F provides a three-dimensional representation of the sliding regions, identifying rows (in red) of high-LENS atoms corresponding to the dislocation vectors. 

The system herein discussed, used as a proof of concept case, proves the reliability of the \textbf{LEAP} analysis scheme. This example demonstrates how, in a purely agnostic manner and simply relying on the concept of local fluctuations and their spatial and temporal correlations, it is possible to revisit and reconstruct non-trivial complex and collective phenomena such as those underpinning the well-known dislocation motion that control the plastic deformation in metals.

\subsection{Into the physics of complex active matter systems with LEAP}
Lastly, we show how a \textbf{LEAP} analysis can provide insights into the physics of complex active matter systems whose trajectory may be obtained experimentally rather than by MD simulations.

\begin{figure*}
 \centering
 \includegraphics[width=16cm]{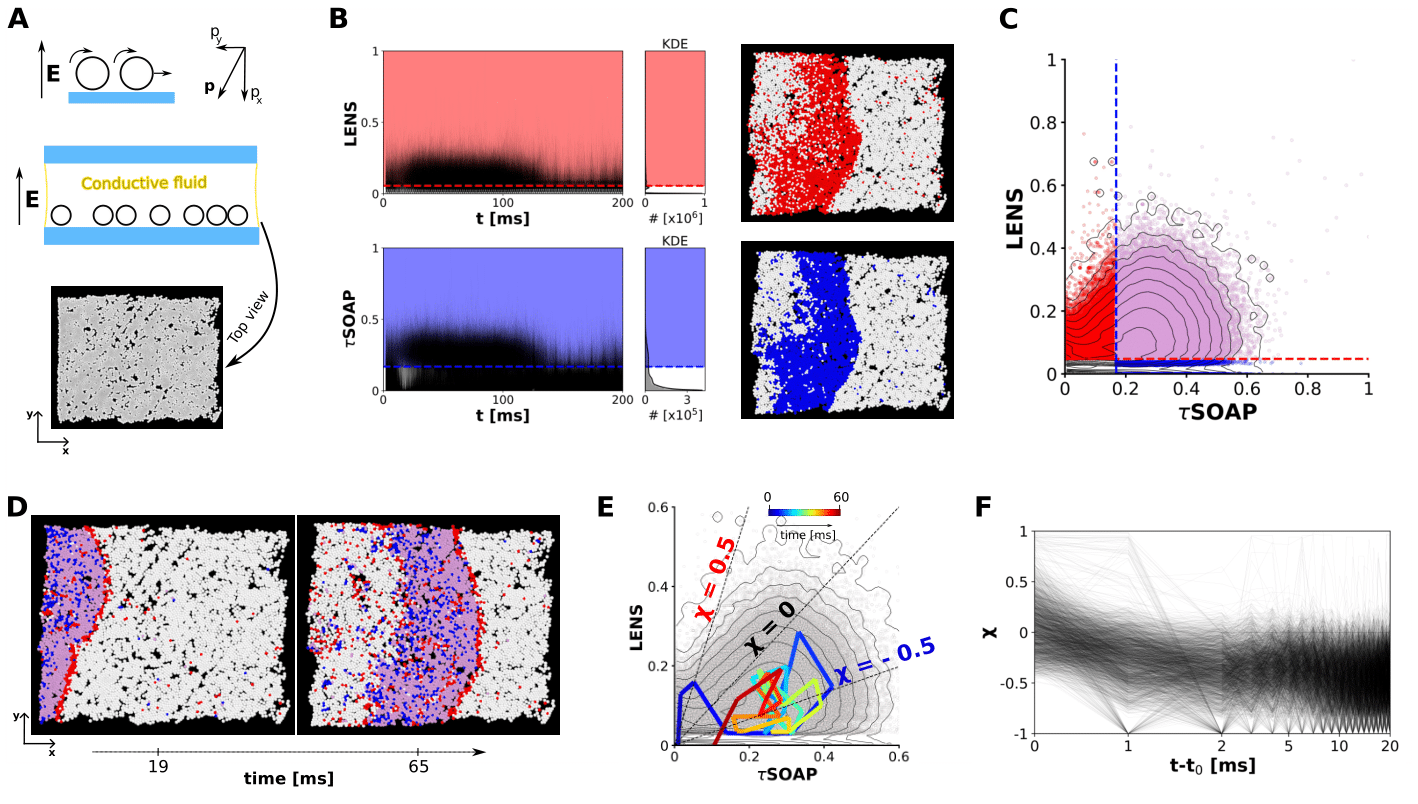}
  \caption{\footnotesize\textbf{\textbf{LEAP} analysis of a microscopic experimental system: Quincke rollers.} (A) Top: Schematic illustration of Quincke rollers, \textit{i.e.} dielectric colloidal particles in a weakly conductive fluid and exposed to a vertical DC electric field, \textbf{E}. Bottom: Top view of the experimental system, extracted from an optical microscope movie.\cite{liu2021activity} N = 6921 particles are tracked for 312 ms, with a sampling time of 1 ms.\cite{becchi2024layer} (B) LENS and \textit{$\tau$}SOAP time-series related to all the tracked particles, with the corresponding KDEs. The first 200 ms are considered. The tails of both the KDEs are isolated: LENS threshold = 0.056 (red dashed line), \textit{$\tau$}SOAP threshold = 0.169 (blue dashed line). They allow visualizing the collective wave flowing left-to-right throughout the system, shown in red for LENS (Top-Right) and in blue for \textit{$\tau$}SOAP (Bottom-Right) in the two example snapshots (t = 50 ms). (C) Projection of the \textbf{LEAP} dataset on the 2D LENS-\textit{$\tau$}SOAP phase space (6921 particles for 200 time steps, for a total of $\sim$ \num{1.3e6} data points). Based on the thresholds reported in (B), four different domains can be distinguished (white, blue, red and lilac). (D) Two example snapshots are colored according to the identified domains shown in (C). (E) Representative example trajectory (particle ID 106) following a characteristic path on the \textbf{LEAP} 2D phase space. 60 ms of the trajectory are shown, colored from blue to red as time increases. (F) $\chi$ parameter related to all the particles in the system remaining in the core of the wave (lilac domain in (C)) for at least 20 ms (2680 particles in total). $t_{0}$ corresponds to the first instant in which every particle enters the red region in (C), that is, the front of the wave.}
\label{fig:fig06}
\end{figure*}
 
As a case study, we use Quincke rollers: Dielectric colloidal (polystyrene) particles immersed in a conducting fluid subjected to an external vertical DC electric field (\textbf{E}), as illustrated in Fig.~\ref{fig:fig06}A.\cite{liu2021activity} In this system, an increase of \textbf{E} enhances the fluctuations of particle polarization vectors \textbf{P}, promoting particles rolling on the \textit{xy} plane. Recent studies have demonstrated how interesting collective phenomena, such as activity waves, vortices, \textit{etc.}, may emerge and propagate in a quiescent population of such colloidal particles in conditions where \textbf{E} is weaker than $\textbf{E}_{c}$, the threshold value promoting the motion of an isolated particle.\cite{liu2021activity} 
Here, we consider the optical microscope movie reported by Liu \textit{et al.},\cite{liu2021activity} where an activity wave emerges and flows directionally, through a field of view of 700 × 700 $\mu$m\textsuperscript{2} containing N = 6921 particles over 0.25 s of image acquisition. From this movie, we obtain a trajectory of T = 312 frames using the Python package Trackpy (see snapshot in Fig.~\ref{fig:fig06}A).\cite{allan2024} Of these, we retain only the first 200 frames in the analysis (the portion of the movie in which the wave passage is observed). From the IDs' positions along the trajectory, we compute the LENS and \textit{$\tau$}SOAP time-series for each particle, as plotted in Fig.~\ref{fig:fig06}B (complete technical details are provided in \textit{Materials and Methods}). 

In both time-series, the high-density peaks in the KDEs can be isolated from the low-density high-intensity domains depicted by the red and the blue regions of Fig.~\ref{fig:fig06}B (see SI Appendix, Fig. S7 for technical details). The high-density peaks, characterized by low-intensity signals, are related to the characteristic diffusive (Top) and structural (Bottom) vibrations of the particles in the system, which represent the "background" LENS and \textit{$\tau$}SOAP noise, respectively. Conversely, the low-density domains of the time-series identify fewer particles undergoing intense LENS or \textit{$\tau$}SOAP fluctuations in the system (high-intensity signals). Coloring these LENS/\textit{$\tau$}SOAP-fluctuating particles in each frame (in red and blue, respectively) allows visualizing the collective wave flowing left-to-right throughout the system (Fig.~\ref{fig:fig06}B: right). Both LENS and \textit{$\tau$}SOAP components capture the wave propagation, but not exactly in an identical manner. 

The projection of the \textbf{LEAP} dataset on the 2D LENS-\textit{$\tau$}SOAP phase space (6921 particles for 200 frames, for a total of $\sim$ \num{1.3e6} data points, Fig.~\ref{fig:fig06}C) allows to distinguish four different domains based on the thresholds discriminating the relevant fluctuations from noise in the LENS and \textit{$\tau$}SOAP dimensions (clustering details in SI Appendix, Fig. S7). Coloring the trajectory frames of Fig.~\ref{fig:fig06}D according to such a classification shows the quiescent particles in the system as depicted in white (corresponding to the bottom-left region of Fig.~\ref{fig:fig06}C: LENS $\leqslant$ 0.056 and \textit{$\tau$}SOAP $\leqslant$ 0.169). These are particles that just vibrate, whose motion constitutes the intrinsic noise in the system. The lilac domain, which includes all the particles in the core of the wave, corresponds to the region near the diagonal in the LENS-\textit{$\tau$}SOAP phase space of Fig.~\ref{fig:fig06}C (LENS $> 0.056$ and \textit{$\tau$}SOAP $> 0.169$). In this region, every particle undergoing a LENS fluctuation is also undergoing a \textit{$\tau$}SOAP fluctuation (structural+diffusive fluctuations). The two most interesting domains are those colored in red (LENS $> 0.056$ and \textit{$\tau$}SOAP $\leqslant$ 0.169) and blue (LENS $\leqslant$ 0.056 and \textit{$\tau$}SOAP $> 0.169$), which identify particles undergoing fluctuations that are LENS- and \textit{$\tau$}SOAP-dominated, respectively. In particular, the red cluster (LENS fluctuations) corresponds to the wavefront (the wave moves left-to-right, see also Movie S2), a region where LENS-dominated fluctuations appear as simultaneous in time and correlated in space. This indicates that the first event that anticipates the passage of the wave in this system is a LENS event (namely, every particle that becomes part of the wave, first undergoes a LENS fluctuation).

We can follow the trajectory of every particle in the system over time on the LENS-\textit{$\tau$}SOAP phase space. Fig.~\ref{fig:fig06}E shows one prototypical trajectory of one example particle (ID 106): Very similar paths are observed for the other particles in the system. In general, the trajectories follow a characteristic path on the 2D \textbf{LEAP} phase space, where the particles move sequentially red $\xrightarrow{}$ blue $\xrightarrow{}$ lilac (where they spend $\sim$20-40 ms) $\xrightarrow{}$ blue $\xrightarrow{}$ white (where they return to a quiescent configuration). In this system, where the raw trajectory obtained from the movie can be sampled at most every 1 ms,\cite{liu2021activity} the first LENS events (\textit{i.e.} the passage through the red region) are very short-lived and last at most $\sim$ 1-2 ms.
For every particle in the system, we calculated the $\chi$ parameter (as defined in  Eq.~\ref{eq:one}) over time. 
Fig.~\ref{fig:fig06}F reports the $\chi$ parameter over time for all the particles in the system that remain in the core of the wave (lilac region) for at least 20 ms. 
All trajectories are phased in such a way that the point in $t-t_{0}=0$ (in the \textit{x}-axis) corresponds to the instant in which every particle undergoes the first LENS event, entering the front of the wave.
The data shows how for all particles entering the wave, the first LENS-dominated fluctuations annihilate within $\sim$3 ms (the majority of them, within 1 ms). Only sparse new LENS fluctuations are seen to emerge after 5 ms of observation. These identify the sparse, local (non-spatially correlated) LENS fluctuations that may emerge in the body of the wave (red particles within the lilac domain in the snapshots of Fig.~\ref{fig:fig06}D).
Interestingly, when the LENS fluctuations annihilate, every particle undergoes a sharp \textit{$\tau$}SOAP fluctuation ($\chi$ $\sim$ -1). This demonstrates how all the particles in the system follow a red $\xrightarrow{}$ blue $\xrightarrow{}$ lilac pathway (like the trajectory of ID: 106 shown in Fig.~\ref{fig:fig06}E).

These results provide a demonstrative example of how such abstract detection-classification-correlation of local fluctuations \textbf{LEAP} approach can provide precious insights into the behavior and the internal phenomena occurring in complex dynamical systems whose physics is unknown \textit{a priori}.  

\subsection*{Conclusions}
In this paper, we demonstrate how the combination of complementary general descriptors allows elucidating the physical behavior of diverse complex systems by leveraging the concept of local fluctuations. In the philosophy of monitoring individuals along the trajectory, \textbf{LEAP} incorporates two complementary fingerprints, LENS and \textit{$\tau$}SOAP, which detect, respectively, permutationally-variant/structurally-invariant (\textit{diffusive}) and permutationally-invariant/structurally-variant (\textit{structural}) local events (Fig.~\ref{fig:fig01}). Essentially, they are fluctuations that typically characterize the local neighborhoods in the trajectory of complex systems. 

Trajectories linearly moving on the 2D \textbf{LEAP} phase space identify molecules (or, in general, units) whose neighbors undergo local structural rearrangements while changing, at the same time, their local dynamics (in terms of dynamical reshuffling). This is the case, for instance, of phase transitions in the MD simulation of ice/liquid water at phase coexistence (Fig.~\ref{fig:fig02}), where structural and diffusive local events simultaneously occur. However, some physical phenomena may be governed by predominantly diffusive (LENS-dominated) or by predominantly structural (\textit{$\tau$}SOAP-dominated) fluctuations, namely by local dynamical events that can be captured exclusively by the LENS or \textit{$\tau$}SOAP component of our \textbf{LEAP} bivariate time-series. Interestingly, such phenomena can be described by strongly non-linear trajectory paths on the 2D \textbf{LEAP} phase space, thus identifying non-trivial dynamical events (from Fig.~\ref{fig:fig03} to Fig.~\ref{fig:fig06}). In particular, LENS-dominated fluctuations identify "rigid" diffusive dynamical events, where units slide while changing, at every time step, the identity of their closest neighbors which, nonetheless, preserve their structural order. On the other hand, \textit{$\tau$}SOAP-dominated fluctuations identify events where units do not dynamically reshuffle within close neighborhoods, but structurally rearrange (local rattling events). Leveraging such a classification in different types of fluctuations allows correlating such different events in space and time, obtaining precious information for understanding a variety of phenomena. 
For example, in the case of the atoms sliding on the \textbf{Cu(211)} FCC surface seen in Fig.~\ref{fig:fig03} and Fig.~\ref{fig:fig04}, our data demonstrates how every Cu atom undergoing a surface diffusive LENS event (Fig.~\ref{fig:fig04}B: in red) sees a drastic increase in the frequency of the structural \textit{$\tau$}SOAP fluctuations (blue) preceding such event (Fig.~\ref{fig:fig04}C: frequency augmented by $\sim65 \%$ compared to the average \textit{$\tau$}SOAP fluctuations in the system).  

At the same time, the spatial correlation between the local events is also fundamental. In several cases discussed herein, it is interesting to observe where local fluctuations may occur. Besides the temporal one, a spatial relationship may emerge between diffusive and structural fluctuations, \textit{e.g.} in the case of \textbf{Cu(211)} FCC surface (Fig.~\ref{fig:fig04}D-E). Or even, local fluctuations may occur simultaneously in time and well localized in space. This identifies collective events, as demonstrated by the LENS-dominated fluctuations which describe the sliding of the dislocation planes in metals entering the plastic region (Fig.~\ref{fig:fig05}). Lastly, the active matter system case study proves how such a data-driven approach, based on the very simple and general concept of local events and their correlations, may help in elucidating the behavior of complex systems whose trajectories are experimentally obtained (Fig.~\ref{fig:fig06}). Our \textbf{LEAP} analysis on Quincke rollers, indeed, unveils a well-defined sequential mechanism followed by all the particles involved in the wave passage: First, colloids undergo a local LENS-dominated fluctuation; then, the \textit{$\tau$}SOAP-dominated fluctuations grow in the system; finally, particles return to a quiescent state \textit{via} \textit{$\tau$}SOAP fluctuations (Fig.~\ref{fig:fig06}E-F). This provides relevant insights into the origin and mechanisms underpinning the evolution and annihilation of such a phenomenon. 

With a similar spirit as some causality detection methods,\cite{granger1969investigating, chen2004analyzing, schreiber2000measuring, paluvs2001synchronization, glielmo2022ranking, del2024robust} these results demonstrate how such abstract \textbf{LEAP} analyses can provide crucial insights useful to understand the mechanisms underlying a variety of physical phenomena and to predict their emergence in the system in space and time. 
In general, we envisage that this \textbf{LEAP} analysis approach will constitute a precious tool to explore and understand complex systems whose physics is not known \textit{a priori}, as well as to revisit known physical phenomena under a new light. 

\section{Methods}
\subsection{Trajectories and pre-processing}
Complete data and details concerning all molecular models and simulation parameters used to get the MD trajectories, as well as the complete \textbf{LEAP} analysis code, are available at: \url{https://github.com/GMPavanLab/LEAP} (that will be replaced with a definitive Zenodo archive upon acceptance of the final version of this paper). 

All the setup details described in the following have been summarized in the SI Appendix, Table S1.

\textbf{Ice/liquid water phase coexistence.}
The atomistic ice/liquid water phase coexistence system is simulated employing the direct coexistence approach at the solid/liquid transition temperature, as recently used.\cite{caruso2023timesoap, crippa2023detecting, crippa2023machine} The \textbf{TIP4P/ICE} water model\cite{abascal2005potential} is chosen to model both the ice $I_{h}$ and the liquid water phase. In the coexistence model, the two phases are put in contact in the same simulation box. In the initial configuration, half of the water molecules (N = 1024) are in the solid ice phase, the other half (N = 1024) in the liquid phase. The MD trajectory analyzed herein is obtained by simulating the system at constant pressure (1 atm) in correspondence of the solid/liquid transition temperature for the employed water model (T = 267.5 K).\cite{garcia2006melting} The GROMACS software is used.\cite{hess2008gromacs} A 100 ns-long MD production run is performed using the same setup of Ref. \cite{caruso2023timesoap} and a sampling time interval of $\Delta$t = 0.001 ns. The results reported in Fig.~\ref{fig:fig02} are obtained by extracting, from the 100 ns-long simulation, 3 ns (from 95 ns to 98 ns, for a total of 3000 frames). Firstly, the MD trajectory has been pre-processed by considering only the oxygen atoms (OW) of the water molecules in the system as representative centers in the \textbf{LEAP} analysis, \textit{i.e.} for the calculation of LENS\cite{crippa2023detecting} and \textit{$\tau$}SOAP\cite{caruso2023timesoap} time-series. For each of the 2048 oxygen atoms in the system, LENS and \textit{$\tau$}SOAP are computed on the sampled MD configurations using a cutoff of 6 \r{A} (second minimum of the OW-OW radial distribution function, thus enclosing the first two solvation shells) and smoothed using a moving average with width = 0.2 ns (200 frames). This analysis has been repeated also analyzing the whole second half of the same 100 ns-long MD trajectory, \textit{i.e.} from 50 to 100 ns, using a sampling time interval of $\Delta$t = 0.1 ns (for a total of 500 frames) and a smoothing width = 8 ns (80 frames), as reported in SI Appendix, Fig. S2.  

\textbf{FCC Cu(211) surface.}
The atomistic model of copper FCC surface \textbf{Cu(211)} studied herein is composed of N = 2400 atoms arranged in 8 layers that model a portion of an infinite surface through periodic boundary conditions. A 150 ns-long MD trajectory is obtained through the development of a deep-potential MD simulation of the \textbf{Cu(211)} surface built training a Neural Network with the DeepMD platform\cite{wang2018deepmd} on DFT data (for details see Ref. \cite{cioni2023innate}). The trajectory is simulated at T = 600 K and with a sampling time interval of $\Delta$t = 6 ps using the LAMMPS software.\cite{thompson2022}. For the analysis reported in Fig.~\ref{fig:fig03} and Fig.~\ref{fig:fig04}, 24 ns are considered (96-120 ns, for a total of 2000 frames extracted every $\Delta$t = 12 ps along the MD trajectory). The analysis is conducted on the three top-most layers (995 atoms), since we are interested in the dynamics of the surface and most of the bulk remains substantially immobile during the MD simulation.\cite{cioni2023innate} For each of the 995 surface Cu atoms, a cutoff of 6 \r{A} is used to compute the LENS and \textit{$\tau$}SOAP descriptors using a $\Delta$t time lapse of 12 ps. Both signals are smoothed by a moving average, using a width = 120 ps (10 frames). 

\textbf{Plastic deformation of bulk Cu.}
The third system studied is a bulk of copper (Cu) FCC crystal composed of 2744 atoms (see Fig.~\ref{fig:fig05}) and simulated at T = 300 K. The MD simulations are conducted with the LAMMPS software package\cite{thompson2022} using an Embedded Atom Method (EAM) potential specifically designed for copper.\cite{mishin2001structural} The MD trajectories are saved using a sampling time interval of $\Delta$t = 5 ps. This potential has been proved to be reliable in mimicking the phenomena occurring within these materials during a tensile fracture test, allowing to estimate a Young modulus which is consistent with the experiments.\cite{rassoulinejad2016evaluation} The system undergoes minimization and then equilibration at 300 K for 2 ns, employing a Nose-Hoover thermostat and barostat to control the temperature and pressure.\cite{evans1985nose}
Following equilibration, a uniaxial tensile deformation is applied along the \textit{x}-direction with a strain rate of 0.1\% strain per ns. The pressure in the \textit{y}- and \textit{z}-directions is maintained at zero during this deformation. The deformation simulation has been carried out for a total of 150 ns. The \textbf{LEAP} analysis is conducted for the entire simulation duration. However, due to the nature of the specific events being investigated, a brief time interval of 0.1 ns (from 138.85 ns to 138.95 ns) is reported in Fig.~\ref{fig:fig04}, showing details of the signals corresponding to the raw (\textbf{LEAP} and LENS) data. A cutoff of 8 \r{A} is used in the \textbf{LEAP} analysis, which has been found as the best compromise between computational efficiency and the amount of information retained.

\textbf{Experimental Quincke rollers trajectory.}
The Quincke rollers trajectory analyzed (in Fig.~\ref{fig:fig06}) is obtained from an optical microscope movie published by Liu \textit{et al.}.\cite{liu2021activity} By means of image recognition and the Trackpy tracking code,\cite{allan2024} the \textit{x} and \textit{y} coordinates related to 6921 colloidal particles for 312 consecutive frames have been obtained, as described in detail in Ref. \cite{becchi2024layer}. Since the collective propagation occurs in the first part of the trajectory, in the analysis we considered the first 200 consecutive frames (0-200 ms). For each particle in the system, LENS and \textit{$\tau$}SOAP are computed using a cutoff of 40 pixel ($\sim$ 56 $\mu$, that takes into account the neighborhood at least up to the third minimum of the radial distribution function). A moving average is used to smooth the signal with width = 2 ms (2 frames).

\subsection{LENS and \textit{$\tau$}SOAP Data Analysis}
For each individual representative center \textit{i} in each system, the LENS\cite{crippa2023detecting} and \textit{$\tau$}SOAP\cite{caruso2023timesoap} values are computed over time along the trajectory. 

As detailed in Ref. \cite{crippa2023detecting}, being $C^{t}_{i}$ an array containing all the individual identities (IDs) of the particles/individuals surrounding the center \textit{i} within a sphere of radius $r_{cut}$ at the time step t, the LENS value, indicated by $\delta_{i}$, is defined as 
\begin{equation}
\centering
    \delta_i^{t+\Delta t}=\frac{\#(C_i^{t} \bigcup C_i^{t+\Delta t} - C_i^{t} \bigcap C_i^{t+\Delta t})}{\#(C_i^t+C_i^{t+\Delta t})},
\label{eq:two}
\end{equation}
where $\Delta{t}$ is the time interval between two consecutive sampled time steps. The first and the second terms of the numerator are the mathematical union and intersection, respectively, of the neighbor IDs within $r_{cut}$ from the center \textit{i} at the time t and t + $\Delta{t}$. Thus, for each individual center \textit{i}, $\delta_{i}$(t) monitors the \textit{i}-th local environment changes in terms of neighbor identities/individuals along the trajectory, ranging from 0 to 1 for persistent to highly dynamic neighborhoods, respectively.

Indicated by $\lambda_{i}$, the instantaneous \textit{$\tau$}SOAP value is defined as  
\begin{equation}
\centering
   \lambda^{t+ \Delta t}_{i} \propto \sqrt{2-2\textbf{p}^{t}_{i}\textbf{p}^{t+ \Delta t}_{i}},
\label{eq:three}
\end{equation}
where $\textbf{p}^{t}_{i}$ is the full SOAP feature vector associated to the \textit{i}-th individual center within $r_{cut}$ at the time step t, as described in detail in Ref. \cite{caruso2023timesoap}. In a nutshell, $\lambda_{i}$(t) tracks the variations of the \textit{i}-th SOAP vector over time, that is, to what extent the atomic environment related to each center in the system changes at every consecutive time interval $\Delta{t}$ in terms of SOAP power spectrum. 

For each system's individual center, therefore, two time-series were obtained, LENS ($\delta_{i}$(t)) and \textit{$\tau$}SOAP ($\lambda_{i}$(t)), tracking over time the neighbor list and the structural variations, respectively.   
In order to reduce the noise, both $\delta_{i}$(t) and $\lambda_{i}$(t) time-series are smoothed by using a moving average with different time widths depending on the analyzed system (apart where explicitly stated otherwise). A time width of 200 frames is used for the ice/liquid water phase coexistence. In order to better detect the emergence of rare and often short-time fluctuations, smaller widths are chosen for the other systems. A time width of 10 and 2 frames have been chosen for the \textbf{Cu(211)} surface and the experimental Quincke rollers system, respectively. Thus, smoothed $\langle\delta_{i}(t)\rangle$ and $\langle\lambda_{i}(t)\rangle$ time-series were obtained. No smoothing is applied in the case of bulk Cu during constant strain rate. For the sake of simplicity, we refer to $\langle\delta_{i}(t)\rangle$ as $\delta_{i}$(t) and to $\langle\lambda_{i}(t)\rangle$ as $\lambda_{i}$(t).  

For each trajectory, both $\delta_{i}$(t) and $\lambda_{i}$(t) are first normalized from 0 to 1 and, then, combined in the \textbf{LEAP} bi-variate time-series defined as 
\begin{equation}
\centering
   \textbf{LEAP}_{i}(t) = (\delta_{i}(t), \lambda_{i}(t)).
\label{eq:four}
\end{equation}
$\textbf{LEAP}_{i}$(t), thus, is a bi-component array related to the individual \textit{i}, which keeps track of neighbor's identity changes (first component) and of structural re-arrangement (second component) over time.

\subsection{Characteristic time estimation for LENS and \textit{$\tau$}SOAP fluctuations}
A quantitative characteristic time estimation of the diffusive and structural events occurring in the FCC \textbf{Cu(211)} surface is provided in Fig.~\ref{fig:fig04}C. In detail, for each Cu atom experiencing a \textit{$\tau$}SOAP fluctuation, \textit{i.e.} crossing the \textit{$\tau$}SOAP outlier domain depicted in blue in Fig.~\ref{fig:fig04}B, we compute the mean time interval ($\tau$) between two successive structural events. Thus, we obtain (i) the probability distribution function of the structural fluctuation frequency. For those atoms visiting the LENS outlier domain (Fig.~\ref{fig:fig04}B, red), the \textit{$\tau$}SOAP fluctuations are only considered before the diffusive (LENS) event. Furthermore, (ii) a subset consisting of Cu atoms which visit the LENS outlier domain after undergoing \textit{$\tau$}SOAP fluctuations is also considered. This allows to distinguish $\tau$ in the atoms undergoing LENS fluctuations after several structural rearrangements. The mean time intervals ($\tau$) between consecutive \textit{$\tau$}SOAP events are used to build the cumulative distribution functions (CDFs) $P_{n\geqslant1}$: 
\begin{equation}
\centering
   P_{n\geqslant1} = 1-e^{-\tau/\tau*},
\label{eq:four}
\end{equation}
where $\tau*$ is the characteristic time scale of structural (\textit{$\tau$}SOAP) fluctuations. The obtained CDFs are thus reported Fig.~\ref{fig:fig04}C, on the left (dashed blue line for (i), solid blue line for the subset (ii)). In addition, (iii) the probability distribution of LENS fluctuations occurring in the system is obtained and the related CDF is reported with a red line (Fig.~\ref{fig:fig04}C, Left): In this case, $\tau*$ represents the characteristic time scale of diffusive (LENS) fluctuations. The same approach is used to compute the CDF related to the number of structural fluctuations needed before observing a diffusive event (see Fig.~\ref{fig:fig04}C, Right). All these $P_{n\geqslant1}$ distributions turned out to be well fitted by the typical Poisson distribution expected for rare events. 

\subsection{Data availability}
Details on the molecular models and on the MD simulations, and additional MD data are provided in the Supplementary Information.
Complete details of all molecular models used for the simulations, and of the simulation parameters (input files, etc.), as well as the complete LEAP analysis code, are available at: \url{https://github.com/GMPavanLab/LEAP} (that will be replaced with a definitive Zenodo archive upon acceptance of the final version of this paper).

\subsection{Acknowledgements}
G.M.P. acknowledges the support received by the European Research Council (ERC) under the European Union’s Horizon 2020 research and innovation program (Grant Agreement no. 818776 - DYNAPOL).

\subsection{Competing interests statement}
The authors declare no competing interests.

\bibliography{bibliography}

\end{document}


\maketitle
\newpage

\begin{table}[H]
\begin{adjustwidth}{-1in}{-1in} 
    \centering
    \begin{tabular}{|c|c|c|c|c|c|c|}
    \hline
        \textbf{SYSTEM} & \shortstack{\textbf{Trajectory}\\ \textbf{Length[ns]}} & \shortstack{\textbf{\# of sampled}\\\textbf{frames} }& \shortstack{\textbf{Sampling}\\\textbf{$\Delta t$ [\si{\nano\second}]}} & \shortstack{\textbf{Smoothing}\\ \textbf{window [frames]}} & \textbf{$r_{cut}$} & \shortstack{\textbf{LENS-\textit{$\tau$}SOAP} \\\textbf{center}} \\ \hline
        \shortstack{Ice/liquid water\\ phase coexistence} & 3 & 3000 & 0.001 & 200 & 6 \si{\angstrom} & OW \\\hline
        \shortstack{Cu(211) FCC\\ 600K} & 24 & 2000 & 0.012 & 10 & 6 \si{\angstrom} & Cu \\\hline
        \shortstack{Cu FCC bulk\\ 300K} & 0.1 & 20 & 0.005 & - & 8 \si{\angstrom} & Cu \\\hline
        \shortstack{Quincke rollers} & \num{200e3} & 200 & \num{1e3} & 2 & 56 $\mu$m & C \\\hline
    \end{tabular}
    \caption{Setup details of all the LEAP analyses conducted in this work.}
    \label{tab:tab01}
\end{adjustwidth}
\end{table}

\cleardoublepage
\FloatBarrier

\begin{figure}[!ht]
 \includegraphics[width=\columnwidth]{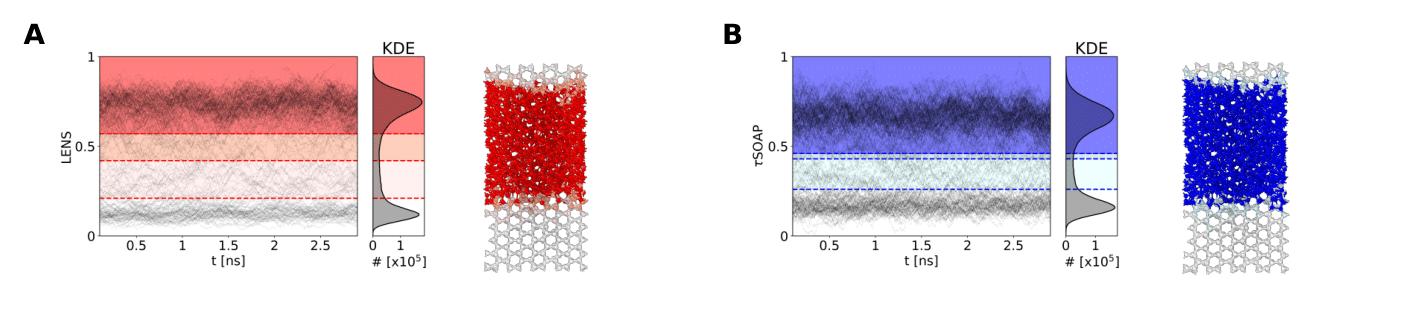}
 \centering
  \caption{\footnotesize\textbf{Univariate Onion clustering analysis on LENS and \textit{$\tau$}SOAP time-series in Ice/liquid water phase coexistence.} LENS (A) and \textit{$\tau$}SOAP (B) time-series, with the corresponding Kernel Density Estimation (KDE) distributions, are computed, for each oxygen (OW), in a system containing 2048 \textbf{TIP4P/Ice} water molecules (half in the crystalline hexagonal ice phase, half in the liquid phase) and simulated at T=267.5 K, namely at the solid/liquid transition temperature for the employed model. The analysis is related to 3 ns extracted from the last part of a 100 ns-long MD simulation, sampled every $\Delta$t=0.001 ns. Both LENS and \textit{$\tau$}SOAP signals are smoothed using a moving average with width=0.2 ns (200 frames). Onion clustering is applied using a time resolution of 0.25 ns (250 frames). (A) Left: Univariate Onion clustering applied on the LENS time-series. Four clusters are detected (LENS thresholds=[0.21, 0.42, 0.57]), colored from white to red. Right: Molecular Dynamics (MD) snapshots colored according to the detected clusters on the left, using the same color code. Four distinct domains are captured: Ice (white), solid/liquid interfaces (mistyrose and lightsalmon, respectively), and liquid water (red). (B) Left: Univariate Onion clustering applied on the \textit{$\tau$}SOAP time-series. Four clusters are captured (\textit{$\tau$}SOAP thresholds=[0.26, 0.43, 0.46]) and colored from white to blue. Right: MD snapshot, colored according to the detected clusters, is shown. Four distinct domains can be observed: Ice (white), solid/liquid interfaces (lightcyan and lightblue, respectively), and liquid water (blue). Both snapshots are taken at t$\sim$1.2 ns.}
   \label{SIfig01}
\end{figure}

\begin{figure}[!ht]
 \includegraphics[width=\columnwidth]{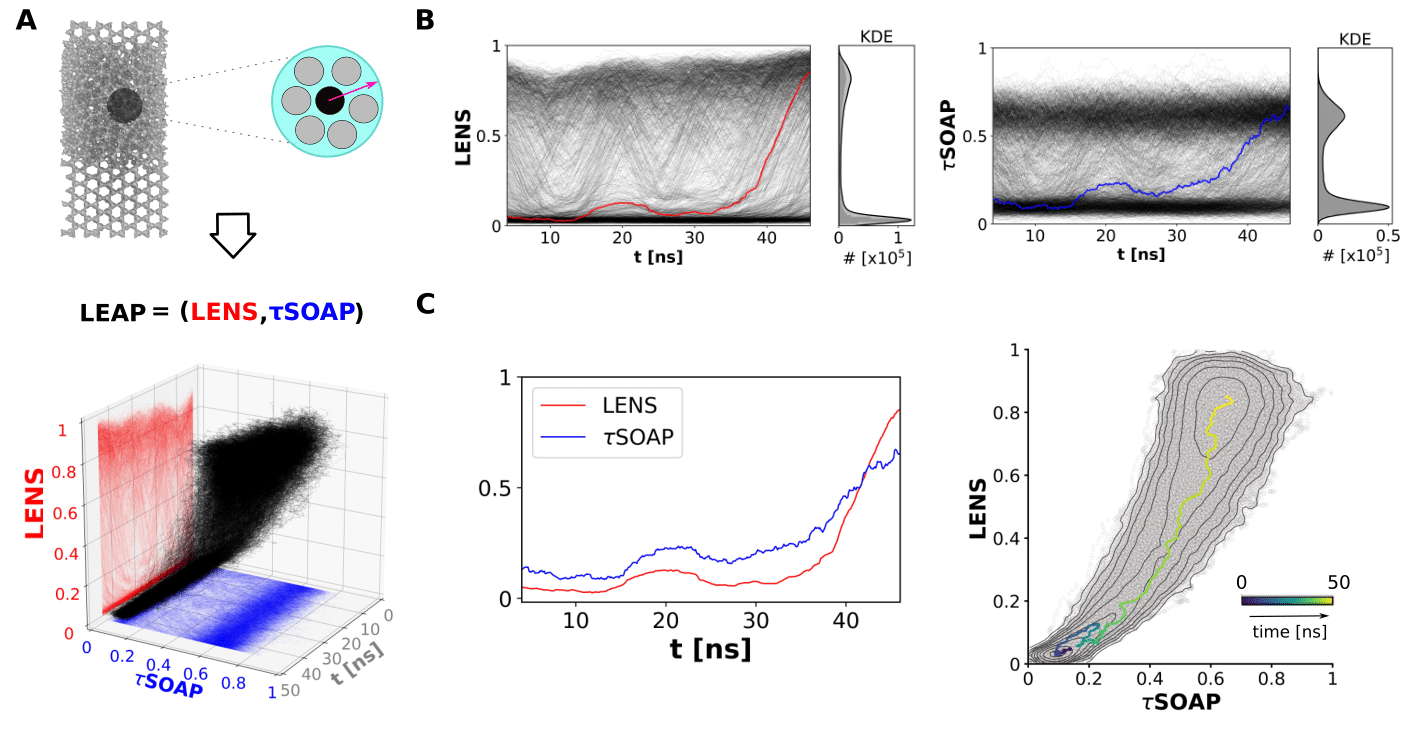}
 \centering
  \caption{\footnotesize\textbf{LEAP analysis in Ice/liquid water phase coexistence extended to 50 ns of a 100 ns-long MD trajectory.} (A) \textbf{LEAP} time-series dataset related to the whole second half of a 100 ns-long MD trajectory (sampling time $\Delta$t = 0.1 ns) composed of 2048 water molecules (\textbf{TIP4P/Ice} water model, whose 50\% in the crystalline hexagonal ice configuration and the remaining 50\% in the liquid phase, coexisting in a dynamic equilibrium. (B) LENS and \textit{$\tau$}SOAP time-series, with the related Kernel Density Estimation (KDE) distributions, for all the water molecules (oxygen centers) in the system. Both time-series are smoothed using a moving average with width=8 ns (80 frames). Signals related to an example water molecule (ID 575) are highlighted on both LENS (red) and \textit{$\tau$}SOAP (blue) components. (C) Left: Plot showing LENS and \textit{$\tau$}SOAP signals related to the ID 575, while simultaneously passing from the low-intensity to the high-intensity KDE peak domain displayed in (B). Right: Projection of the whole \textbf{LEAP} dataset on the 2D LENS-\textit{$\tau$}SOAP phase space (2048 water molecules x 500 frames, for a total of $\sim$\num{1e6} data points). The \textbf{LEAP} path related to the ID 575 is colored from blue to yellow as time increases.}
   \label{SIfig02}
\end{figure}

\begin{figure}[!ht]
 \includegraphics[width=\columnwidth]{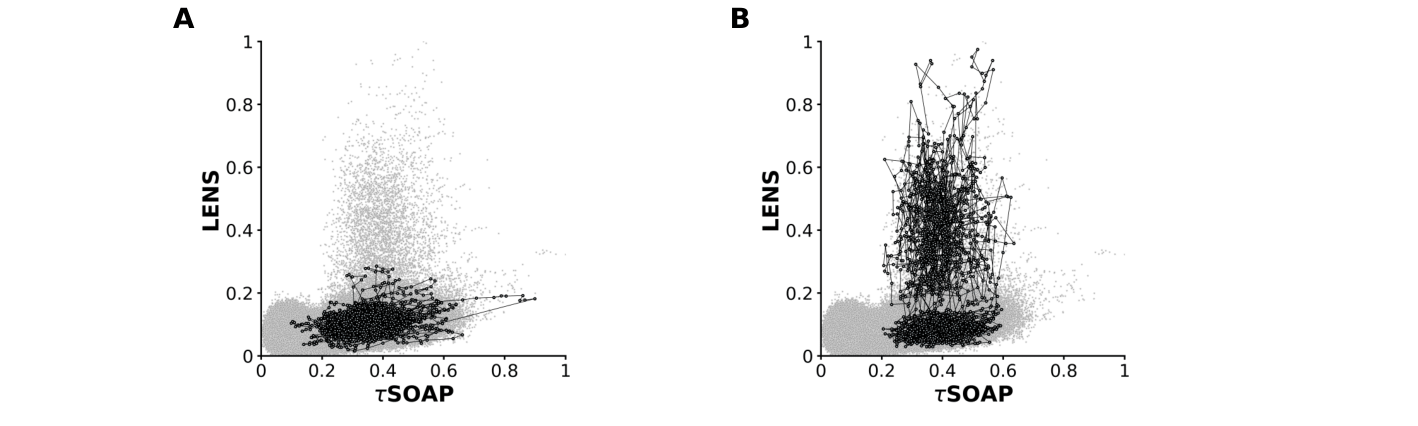}
 \centering
  \caption{\footnotesize\textbf{Examples of trajectory paths projected on the 2D \textbf{LEAP} phase space related to the Cu(211) surface at T = 600 K.} The \textbf{LEAP} dataset (related to the Cu atoms belonging to the three top-most surface layers -995 atoms- at each time step -2000 frames-, for a total of $\sim$\num{2e6} data points) is projected on the 2D LENS-\textit{$\tau$}SOAP phase space, displayed in gray. On such plot, two different example trajectory paths are highlighted, respectively, in (A) and (B). (A) The trajectory path related to the ID 170 is reported (black), which explores the phase space mainly moving along the \textit{$\tau$}SOAP dimension. (B) The trajectory path related to the ID 59 is shown (black), which moves along the LENS dimension. Therefore, we report in (A) a first representative atom experiencing predominantly \textit{$\tau$}SOAP (\textit{structural}) fluctuations, while in (B) a second representative atom experiencing predominantly LENS (\textit{diffusive}) fluctuations.}
   \label{SIfig03}
\end{figure}

\begin{figure}[!ht]
 \includegraphics[width=\columnwidth]{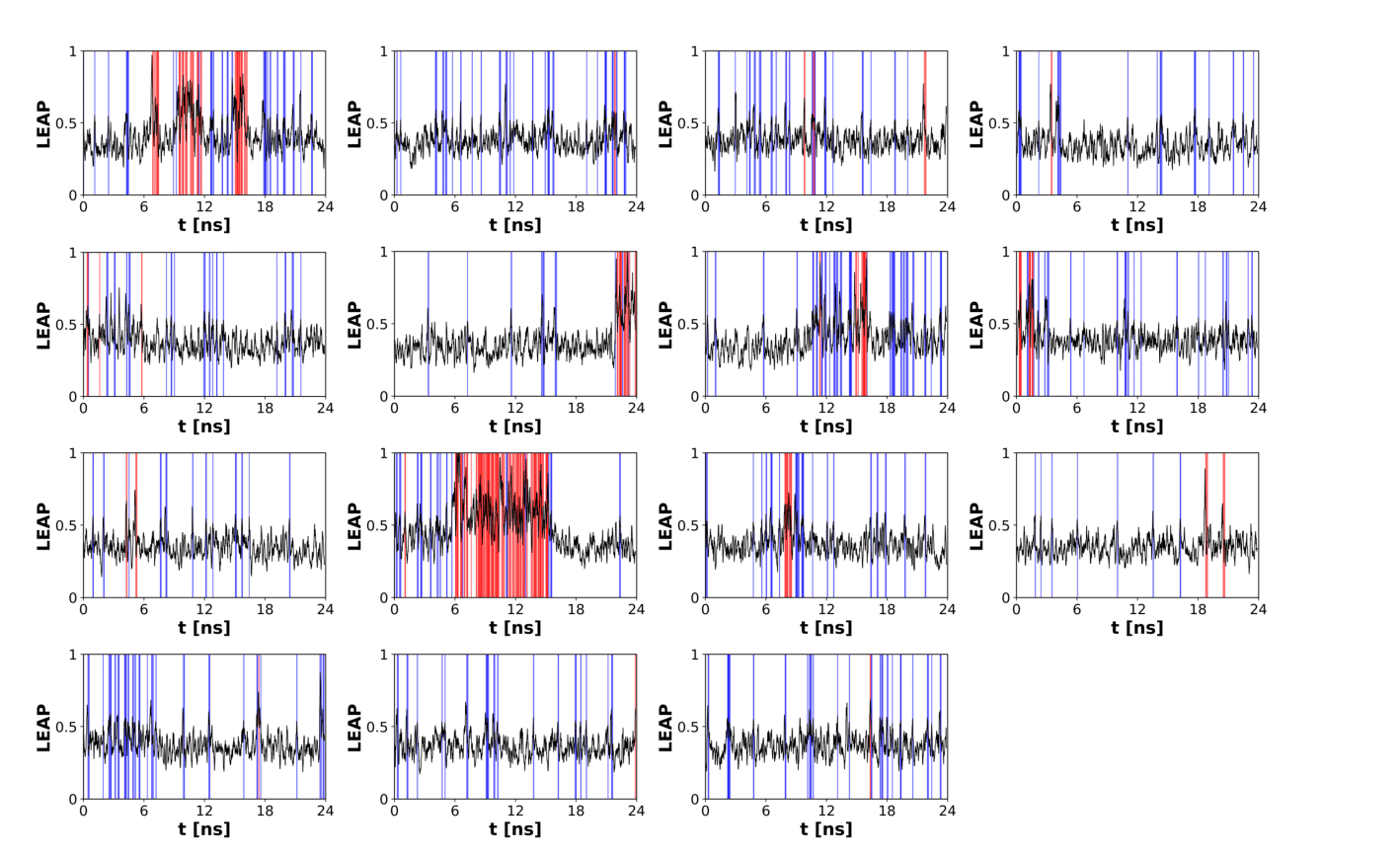}
  \centering
   \caption{\footnotesize\textbf{\textbf{LEAP} time-series of atoms experiencing diffusive fluctuations on the Cu(211) surface at T = 600 K.} Shown as black signals, \textbf{LEAP} (magnitude) time-series related to the atoms experiencing, along the trajectory, at least one predominantly LENS (\textit{diffusive}) fluctuation. On the basis of the dynamical event they are experiencing (predominantly \textit{$\tau$}SOAP or predominantly LENS), and by exploiting the classification reported in the main paper (Fig. 4), the transit through LENS and \textit{$\tau$}SOAP outlier domains are identified with red and blue bands, respectively.}
   \label{SIfig04}
\end{figure}

\begin{figure}[!ht]
 \includegraphics[width=\columnwidth]{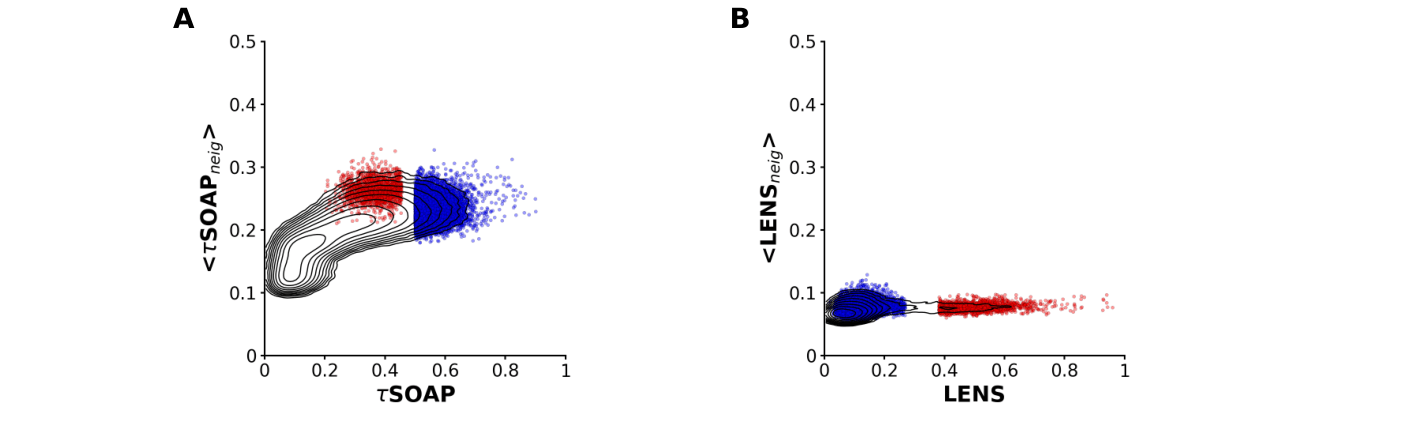}
 \centering
   \caption{\footnotesize\textbf{Spatial relationships in the Cu(211) surface at T=600 K.}
   (A) ID \textit{$\tau$}SOAP value \textit{vs.} \textit{$\tau$}SOAP mean value of its neighbors. For each ID in each MD time step ($\sim$\num{2e6} data points), the plot shows the relationship between the ID \textit{$\tau$}SOAP value and the \textit{$\tau$}SOAP mean value of its neighbors. (B) ID LENS value \textit{vs.} LENS mean value of its neighbors. For each ID in each MD time step ($\sim$\num{2e6} data points), the plot shows the relationship between the ID LENS value and the LENS mean value of its neighbors. In both plots, the red and blue points are related, respectively, to the LENS and \textit{$\tau$}SOAP fluctuations identified in the main paper (Fig. 4).}
   \label{SIfig05}
\end{figure}

\begin{figure}[!ht]
 \includegraphics[width=\columnwidth]{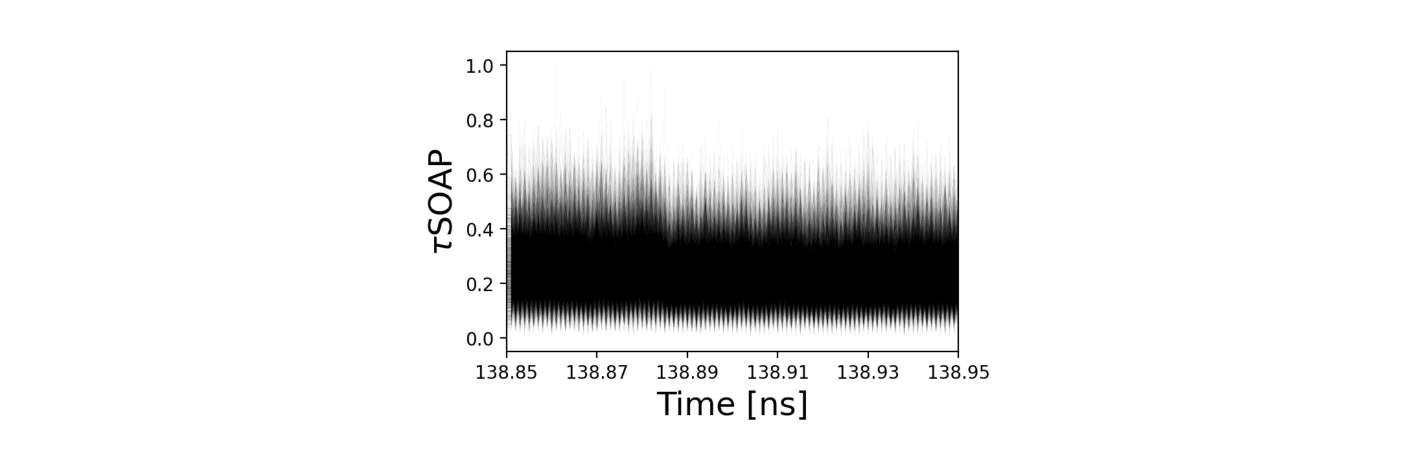}
 \centering
  \caption{\footnotesize\textbf{\textit{$\tau$}SOAP component during the plastic deformation of metals.} Decomposition of the \textbf{LEAP} time-series, related to 2744 atoms in a bulk of copper (Cu) FCC crystal and subjected to a constant strain rate at $T = 300$ K, in its \textit{$\tau$}SOAP component.
   As shown in the main paper (Fig. 5), the reference plastic event occurs at $\sim$ 130-145 ns. A zoom in the temporal evolution of the \textit{$\tau$}SOAP time-series (from 138.85 ns to 138.95 ns) does not reveal any pronounced peak, differently from what observed for the LENS component at $\sim$ 138.88 ns in the main paper.}
   \label{SIfig06}
\end{figure}

\begin{figure}[!ht]
 \includegraphics[width=\columnwidth]{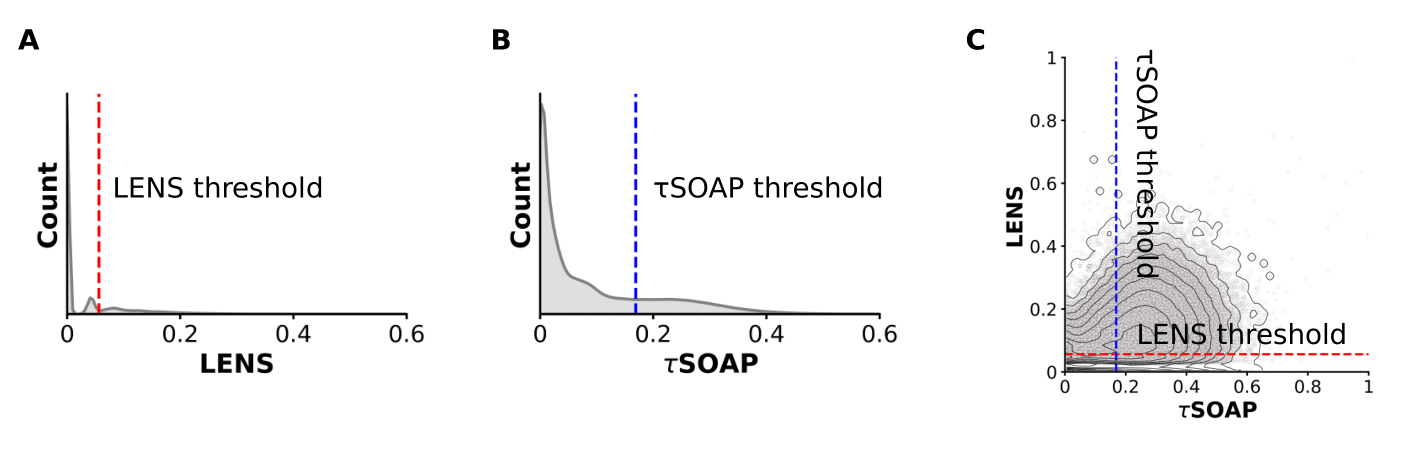}
 \centering
   \caption{\footnotesize\textbf{Identification of LENS and \textit{$\tau$}SOAP fluctuations in the active matter system.} (A) Kernel Density Estimation (KDE) distribution of the LENS time-series computed, at each frame (200 frames in total, corresponding to 200 ms) for each particle (N = 6921). A low-density domain, characterized by fewer particles undergoing high-intensity LENS values, can be isolated by the LENS threshold (LENS = 0.056, red dashed line). The LENS threshold corresponds to the minimum value between the last two peaks identified in the KDE distribution. All the LENS values higher than the threshold are considered as LENS fluctuations. (B) KDE distribution of the \textit{$\tau$}SOAP time-series computed, at each frame (200 frames in total, corresponding to 200 ms) for each particle (N = 6921). Such as in (A) for LENS, the \textit{$\tau$}SOAP threshold (\textit{$\tau$}SOAP = 0.169, blue dashed line) isolates a low-density domain, characterized by high-intensity \textit{$\tau$}SOAP values. The \textit{$\tau$}SOAP threshold corresponds to the minimum value between the last two peaks identified in the KDE distribution. All the \textit{$\tau$}SOAP values higher than the threshold are considered as \textit{$\tau$}SOAP fluctuations. These thresholds are used to separate fluctuations from the characteristic LENS and \textit{$\tau$}SOAP vibrations (low-intensity values) in the system. (C) The thresholds identified in (A) and (B) are plotted on the projection of the \textbf{LEAP} dataset (6921 particles for 200 frames, for a total of $\sim$\num{1.3e6} data points, shown in gray) on the 2D LENS-\textit{$\tau$}SOAP phase space.}
   \label{SIfig07}
\end{figure}

\cleardoublepage
\FloatBarrier